\documentclass[%
 reprint,
 superscriptaddress,
 amsmath,amssymb,
 aps,
]{revtex4-2}

\usepackage{graphicx}
\usepackage{dcolumn}
\usepackage{bm}
\usepackage{hyperref}

\let\vec\mathbf								

\usepackage{textcomp}						
\usepackage{cleveref}						
\usepackage{amsmath,amssymb,amsfonts}		
\usepackage{relsize}						
\usepackage{mathtools}						
\usepackage{chngcntr}
\usepackage{setspace}
\usepackage{xcolor}
\usepackage{dsfont}
\usepackage{graphicx}
\usepackage{mathrsfs}

\crefname{equation}{Eq.}{Eqs.}
\crefname{figure}{Fig.}{Figs.}
\crefname{table}{Tab.}{Tabs.}
\crefname{section}{Sec.}{Secs.}
\crefname{appendix}{Appendix}{Appendices}
\Crefname{equation}{Equation}{Equations}
\Crefname{figure}{Figure}{Figures}
\Crefname{table}{Table}{Tables}
\Crefname{section}{Section}{Sections}

\DeclareMathAlphabet{\mathcal}{OMS}{cmsy}{m}{n}

\usepackage{pxfonts}

\DeclareMathAlphabet{\mathsf}{OT1}{cmss}{m}{n}
\SetMathAlphabet{\mathsf}{bold}{OT1}{cmss}{bx}{n}

\definecolor{darkblue}{rgb}{0,0,.6}
\definecolor{darkred}{rgb}{.6,0,0}
\definecolor{darkgreen}{rgb}{0,.6,0}

\newcommand{\um}{\text{\textmu m}}		
\newcommand{\I}{\operatorname{i}}		
\newcommand{\e}{\operatorname{e}}		
\newcommand*\diff{\mathop{}\!\mathrm{d}} 
\newcommand{\DFT}{\text{FT}}			
\newcommand{\iDFT}{\text{FT}^{-1}}		

\newcommand{\ie}{i.\,e.}				
\newcommand{\eg}{e.\,g.}				
\newcommand{\ea}{\textit{et al.}}		

\newcommand{\dn}{\Delta n}				
\newcommand{\dmyelin}{t_{\text{sheath}}} 
\newcommand{\nmyelin}{n_{\text{m}}}		

\newcommand{\kx}{k_{\text{x}}}			
\newcommand{\ky}{k_{\text{y}}}
\newcommand{\kz}{k_{\text{z}}}
\newcommand{\kxy}{k_{\text{xy}}}

\newcommand{\dx}{\Delta x}				
\newcommand{\dy}{\Delta y}				
\newcommand{\dz}{\Delta z}				

\usepackage{tikz}  

\usetikzlibrary{calc,shapes.geometric,arrows}
\tikzstyle{step1} = [rectangle, line width = 1pt, rounded corners, minimum width=3cm, minimum height=1cm,text centered, draw=green, fill=green!5]
\tikzstyle{step2} = [rectangle, line width = 0.8pt, minimum width=3cm, minimum height=1cm, text centered, draw=black, fill=yellow!5]
\tikzstyle{step3} = [rectangle, line width = 1pt, rounded corners, minimum width=3cm, minimum height=1cm, text centered, draw=orange, fill=red!5]
\tikzstyle{step4} = [rectangle, line width = 0.8pt, minimum width=3cm, minimum height=1cm, text centered, draw=black, fill=blue!5]
\tikzstyle{step5} = [rectangle, line width = 0.8pt, minimum width=3cm, minimum height=1cm, text centered, draw=black, fill=orange!5]
\tikzstyle{step6} = [rectangle, line width = 1pt, rounded corners, minimum width=3cm, minimum height=1cm, text centered, draw=red, fill=red!5]
\tikzstyle{p} = [circle, line width = 0pt, minimum size = 0pt, inner sep = 0pt, fill=black]
\tikzstyle{arrow} = [thick,->,>=stealth, line width=1pt]

\begin{document}

\preprint{APS/123-QED}

\title{Finite-Difference Time-Domain simulations of transmission microscopy enable\\ a better interpretation of 3D nerve fiber architectures in the brain}

\author{Miriam Menzel}
 \email{m.menzel@fz-juelich.de}
 \affiliation{Institute of Neuroscience and Medicine (INM-1), Forschungszentrum J\"{u}lich, 52425 J\"{u}lich, Germany}
\author{Markus Axer}
 \affiliation{Institute of Neuroscience and Medicine (INM-1), Forschungszentrum J\"{u}lich, 52425 J\"{u}lich, Germany}
\author{Hans De Raedt}
 \affiliation{Zernike Institute for Advanced Materials, University of Groningen, 9747AG Groningen, the Netherlands}
\author{Irene Costantini}
 \affiliation{National Institute of Optics -- Italian National Research Council (INO-CNR), 50125 Firenze, Italy}
 \affiliation{European Laboratory for Non-Linear Spectroscopy, University of Florence, 50019 Sesto Fiorentino, Italy}
\author{Ludovico Silvestri}
 \affiliation{European Laboratory for Non-Linear Spectroscopy, University of Florence, 50019 Sesto Fiorentino, Italy}
\author{Francesco S.~Pavone}
 \affiliation{National Institute of Optics -- Italian National Research Council (INO-CNR), 50125 Firenze, Italy}
 \altaffiliation[Also at ]{Department of Physics, University of Florence, 50019 Sesto Fiorentino, Italy}
\author{Katrin Amunts}
 \affiliation{Institute of Neuroscience and Medicine (INM-1), Forschungszentrum J\"{u}lich, 52425 J\"{u}lich, Germany}
 \affiliation{C\'{e}cile and Oskar Vogt Institute for Brain Research, University Hospital D\"{u}sseldorf, University of D\"{u}sseldorf, 40204 D\"{u}sseldorf, Germany}
\author{Kristel Michielsen}
 \affiliation{J\"{u}lich Supercomputing Centre, Forschungszentrum J\"{u}lich, 52425 J\"{u}lich, Germany}

\date{\today}


\begin{abstract}
In many laboratories, conventional bright-field transmission microscopes are available to study the structure and organization principles of fibrous tissue samples, but they usually provide only 2D information. To access the third (out-of-plane) dimension, more advanced techniques are employed. An example is 3D Polarized Light Imaging (3D-PLI), which measures the birefringence of histological brain sections to derive the spatial nerve fiber orientations. 
Here, we show how light scattering in transmission microscopy measurements can be leveraged to gain 3D structural information about fibrous tissue samples like brain tissue. For this purpose, we developed a simulation framework using finite-difference time-domain (FDTD) simulations and high performance computing, which can easily be adapted to other microscopy techniques and tissue types with comparable fibrous structures (\eg, muscle fibers, collagen, or artificial fibers). 
As conventional bright-field transmission microscopy provides usually only 2D information about tissue structures, a three-dimensional reconstruction of fibers across several sections is difficult. By combining our simulations with experimental studies, we show that the polarization-independent transmitted light intensity (transmittance) contains 3D information: We demonstrate in several experimental studies on brain sections from different species (rodent, monkey, human) that the transmittance decreases significantly (by more than 50\%) with the increasing out-of-plane angle of the nerve fibers. 
Our FDTD simulations show that this decrease is mainly caused by polarization-independent light scattering in combination with the finite numerical aperture of the imaging system. This allows to use standard transmission microscopy techniques to obtain 3D information about the fiber inclination and to detect steep fibers, without need for additional measurements or changes in the experimental setup. Furthermore, we demonstrate that the transmittance can be used to classify regions with low birefringence signals, like regions with in-plane crossing fibers and regions with out-of-plane fibers, which can to date not be distinguished in 3D-PLI measurements. This enables a much better reconstruction of the complex nerve fiber architecture in the brain.
\end{abstract}

\maketitle


\section{Introduction}
\label{sec:introduction}

The human brain consists of a huge network of nerve fibers: Around 100 billion nerve cells are connected to 10,000 other nerve cells on average \cite{gilbert2000,herculano2009,kandel,squire2008,longstaff2011}. Understanding the structure and function of the brain remains a key challenge for neuroscience. To figure out how brain function emerges from its structural organization, it is necessary to study the neuronal connections, \ie, the three-dimensional nerve fiber architecture of the brain. Developing a detailed network model of the brain, the so-called \textit{connectome} \cite{behrens2012,shi2017}, reveals connected brain regions and helps to identify important nerve fiber connections, which is a prerequisite for brain surgery.
It also serves as a reference for fiber tractography algorithms, improving the interpretation of clinical data obtained from diffusion magnetic resonance imaging (MRI) \cite{mori2006, beaulieu2002, tuch2003}. Finally, the connectivity of the nerve fibers exposes pathological changes in the brain's tissue structure, allowing to study neuro-degenerative diseases like Alzheimer's or Parkinson's disease and to develop new treatments and tools for improved diagnostics.
To visualize and derive brain tissue properties and organization principles, light-microscopy techniques are widely used \cite{golgi1873,cajal1904,nieuwenhuys2012}.

In this paper, we study how the scattering of light can be used to improve the interpretation of light microscopy images from fibrous tissue samples like brain tissue. For this purpose, we have developed a simulation framework that models light scattering in transmission microscopy measurements and allows an improved interpretation of the measured data by delivering 3D information about the underlying fiber structure.

Technological progress and new advances in tissue preparation and labeling have enabled the development of techniques that reveal the 3D nerve fiber architecture in both living and post-mortem brains \cite{osten2013}, 
such as Optical Coherence Tomography \cite{men2016,magnain2015,benArous2011}, Micro-Optical Sectioning Tomography \cite{li2010}, Light-Sheet Microscopy \cite{susaki2014,costantini2015,mertz2010,stefaniuk2016,silvestri2014} or \textit{Two-Photon Fluorescence Microscopy (TPFM)} \cite{wang2013,zong2017,amato2016,kawakami2012, silvestri2014}.
While most of these techniques are limited to small sample sizes, the neuroimaging technique \textit{3D Polarized Light Imaging (3D-PLI)} \cite{MAxer2011_1,MAxer2011_2} allows to study the nerve fiber architecture of whole post-mortem brains with microscopic resolution: unstained histological brain sections are illuminated by polarized light and the birefringence caused by the highly anisotropic structure of the nerve fibers is measured, thus revealing their spatial orientations \cite{menzel2015}.

Such neuroimaging techniques require special instruments that are not available in each laboratory. Many laboratories are equipped with simple transmission microscopes which only extract 2D information of the investigated tissue structures, making it difficult to reconstruct complex nerve fiber architectures across several brain sections.
Here, we show that conventional bright-field transmission microscopy measurements can be used to obtain information about the 3D fiber structure of a sample, without need to change the experimental setup or to repeat measurements: the polarization-independent intensity of light that is transmitted through the sample (\textit{transmittance}) depends on the orientation of the fibers with respect to the light beam, \ie, the transmittance of a brain section provides information about the out-of-plane \textit{inclination} angle of the enclosed nerve fibers. These findings can also be transferred to other biological and non-biological samples with comparable fibrous structures, \eg, muscle fibers, collagen, or artificial fibers.

The correct reconstruction of nerve fiber crossings is a major challenge for many neuroimaging techniques and a prerequisite for the correct interpretation of clinical MRI data, allowing for better diagnostics and treatments of neuro-degenerative diseases.
In standard 3D-PLI measurements, brain regions with in-plane crossing fibers cannot be distinguished from regions with out-of-plane fibers or regions with low fiber densities because they all yield small birefringence signals. So far, only the strength and phase of the birefringence signal are used to derive the spatial fiber orientations. The transmittance, which is the average value of the signal, has not been used for this purpose. Here, we demonstrate that the transmittance cannot only be used to gain information about the out-of-plane inclination of the fibers, but also to classify these regions and to identify crossing fibers that cannot unambiguously be determined with standard polarization microscopy techniques.

The transmittance is a measure of how much the light is attenuated when it passes through the brain tissue, \ie, it depends on tissue absorption as well as scattering of light. As the absorption coefficient of brain matter is small (less than 0.1\,mm$^{-1}$ \cite{schwarzmaier1997,yaroslawsky2002}), the measured transmittance is expected to be mainly influenced by scattering. To study such complex light-tissue interactions at the microscopic level, we employed \textit{finite-difference time-domain (FDTD)} simulations to compute the propagation of the light wave through the brain tissue sample \cite{taflove,menzel2016,wilts2014,DeRaedt2012,wilts2012}. 
FDTD simulations are a proven tool for studying for example light scattering in lithography applications \cite{taflove2013,azpiroz2008,erdmann2006} or nanostructures \cite{mcmahon2010,grimault2006,irannejad2013}. They have also been applied to investigate microscopy measurements of non-biological and biological tissue samples \cite{taflove2013,drezek2000,kosmas2004}, but not yet to brain tissue. One reason might be that simulations of tissue samples with dimensions of several micrometers are computationally very intense because the mesh size in the simulation needs to be much smaller than the wavelength. To still enable the investigation of larger samples like brain tissue, we used high-performance computing and a simplified simulation model for the optics of the imaging system and the inner structure of the nerve fibers.
The developed simulation framework can easily be adapted to microscopy techniques with different optics (wavelength, polarization of light, numerical aperture, etc.) and tissue samples with comparable fibrous structures. 

The paper is divided in an experimental and a simulation part: 
In \cref{sec:exp-studies}, we evaluate experimental data and develop techniques to obtain structural 3D information from transmittance images of samples with unknown substructure: we study measurement results obtained from brain sections of different species (rodent, monkey, human) and show that the transmittance decreases significantly (by more than 50\%) with increasing out-of-plane inclination angle of the nerve fibers, using both 3D-PLI and TPFM measurements to access the fiber inclination. 
In \cref{sec:sim-studies}, we introduce the FDTD simulation framework and present the simulation results for artificial nerve fiber configurations with different inclination and crossing angles. The simulations show that the decrease in transmittance is mainly caused by isotropic light scattering and by the finite numerical aperture of the imaging system. The in-plane crossing angle of the fibers has no impact on the transmittance and can be determined from the respective scattering pattern.
In \cref{sec:exp-sim}, we combine the simulation results with experimental data and show that the transmittance can be used to distinguish between regions with in-plane crossing fibers, regions with out-of-plane fibers, and regions with low fiber density by combining the transmittance and the strength of the measured birefringence signal.


\section{Experimental studies}
\label{sec:exp-studies}

In this section, various experimental studies are presented that evaluate how the transmittance of brain sections depends on the out-of-plane inclination angle of the enclosed nerve fibers. 
For our studies, we mostly used 3D-PLI measurements as they provide both the transmittance and the three-dimensional nerve fiber orientations independently from each other \cite{MAxer2011_1, MAxer2011_2}.
As both scattering and absorption contribute to the attenuation (transmittance) of light, differences in the transmittance might not only be caused by different fiber inclinations, but also by a different tissue composition or density.
To investigate the inclination dependence independently from tissue composition or preparation, we performed our studies on different species, subjects, and brain sections.

The studies were conducted on healthy brains from mice, rats, vervet monkeys, and humans. All brains were obtained directly after death in accordance with legal and ethical requirements. The brains were deeply frozen, cut into sections of 60\,\textmu m thickness, embedded in a solution of 20\% glycerin, and cover-slipped. 
A detailed description of the brain preparation can be found in \cref{sec:brain-preparation}.
The brain sections were measured with 3D-PLI, using a polarimeter with a numerical aperture of 0.15 and an object-space resolution of about 1.33\,\um\,/\,px \cite{MAxer2011_1,MAxer2011_2}. The polarimeter consists of an LED light source, a rotating linear polarizer, a specimen stage containing the brain section, a circular polarization analyzer, and a camera which records the transmitted light intensity for different rotation angles $\{0^{\circ}, 10^{\circ}, \dots, 170^{\circ}\}$ of the polarizer. More information about the 3D-PLI measurement can be found in \cref{sec:3DPLI}. The signals provide information about the spatial orientations of the nerve fibers.
The amplitude of the signal (\textit{retardation} $\vert\sin\delta\vert$) is related to the birefringence of the brain section and served here as a measure of the out-of-plane fiber inclination angle $\alpha$, using $\delta \propto \cos^2\alpha$ \cite{menzel2015}. 
The transmittance was computed from the same signal without additional measurements by averaging the measured light intensities over all rotation angles. (For an ideal system, conventional transmission microscopy with unpolarized light should yield the same result.) To only consider effects caused by the brain tissue, the resulting transmittance values were normalized for each image pixel by the average transmitted light intensity without sample,
yielding the \textit{normalized transmittance} $I_{\text{T,N}}$.


\subsection{Transmittance of flat and steep nerve fibers}
\label{sec:transmittance-flat-steep}

\begin{figure}[!t]
	\centering
	\includegraphics[width=\columnwidth]{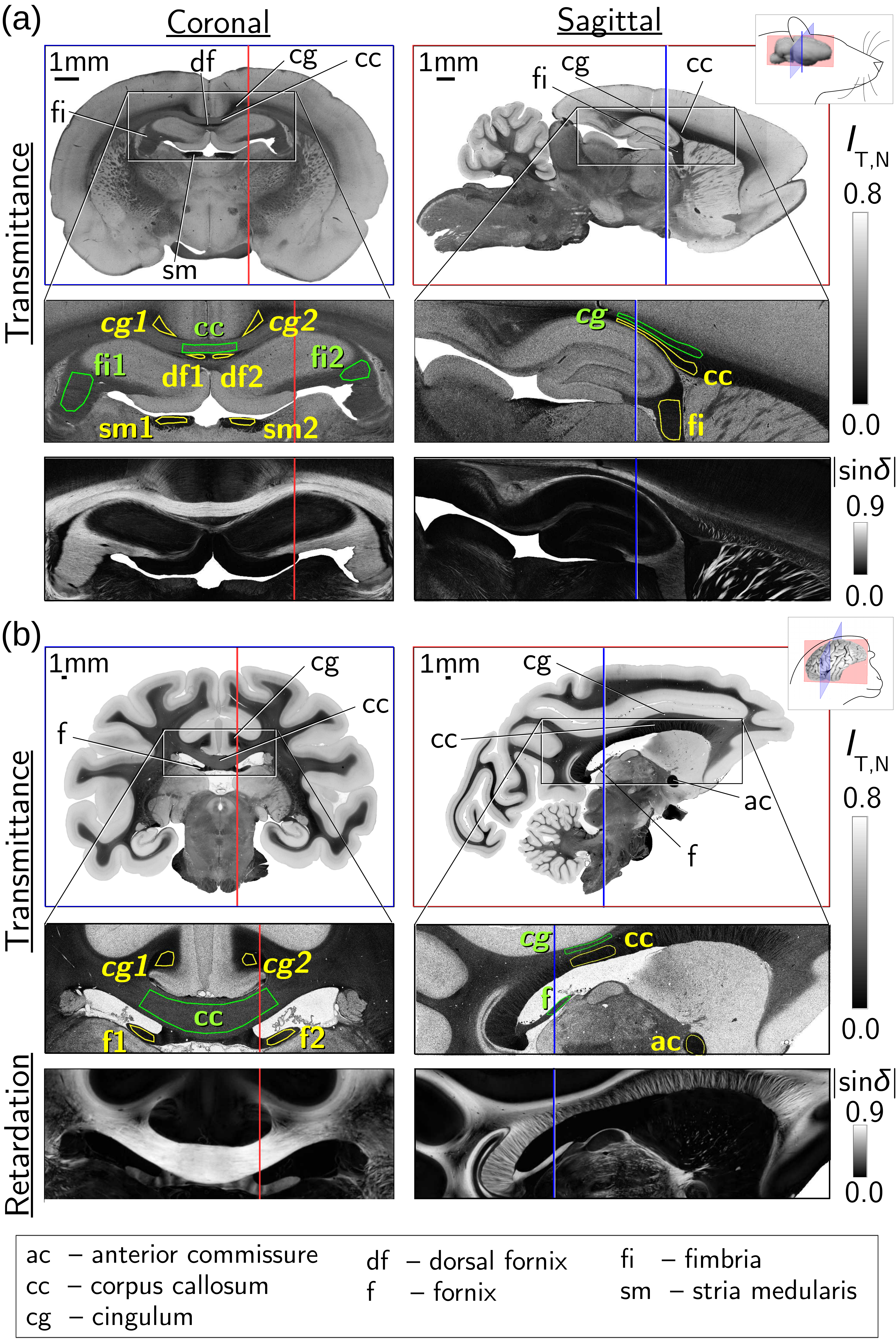}
	\caption{Transmittance and retardation images of coronal and sagittal brain sections for a rat (\textbf{a}) and a vervet monkey (\textbf{b}). The coronal (sagittal) section planes are indicated by blue (red) lines in the respective other brain section for reference, selected anatomical structures are labeled (see legend). The upper two rows of each panel show the normalized transmittance images $I_{\text{T,N}}$, the third row shows the retardation images $\vert\sin\delta\vert$, both obtained from 3D-PLI measurements with 1.33\,\um\ pixel size. The images in the first row show the whole brain section, the images in the second and third row show an enlarged view. The selected regions in yellow (green) belong to steep (flat) nerve fibers, which appear dark (bright) in the transmittance and retardation images.}
	\label{fig:Rat_Vervet_Atlas}
\end{figure}

In order to study how the transmittance depends on the out-of-plane inclination angle of the nerve fibers, the substructure of the investigated brain section needs to be taken into account. 
In large anatomical structures with densely packed nerve fibers, however, the inclination angles of the nerve fibers cannot be exactly determined by 3D-PLI or TPFM measurements. Instead, we compared the transmittance of \textit{flat} nerve fibers (with inclination angles $\alpha < 45^{\circ}$) to the transmittance of \textit{steep} nerve fibers ($\alpha > 45^{\circ}$) by investigating sections from brains that were cut along mutually orthogonal anatomical planes:
one brain was cut along the \textit{coronal} plane (dividing the brain into back and front), the other brain was cut along the \textit{sagittal} plane (dividing the brain into left and right).
As the sagittal plane is oriented perpendicular to the coronal plane, the transmittance of the same brain region can be evaluated for flat nerve fibers in one section plane and for steep nerve fibers in the other section plane. 
Since different brain sections from different specimens might not be comparable due to differences in the tissue structure, we only compared the transmittance values within the same brain section.

\Cref{fig:Rat_Vervet_Atlas} shows the normalized transmittance images $I_{\text{T,N}}$ of coronal and sagittal sections from rat and vervet monkey brains. 
The coronal (sagittal) section planes are indicated by blue (red) lines in the respective other brain section for reference. 
Selected brain structures were identified according to rat \cite{zilles1985,paxinos2007,papp2014} and vervet \cite{woods2011,vervetatlas,zilles2015} brain atlases.
The transmittance values were evaluated in brain regions that have a relatively homogeneous tissue composition and that include predominantly flat nerve fibers (areas surrounded in green) or steep nerve fibers (areas surrounded in yellow). 

The approximate orientation of the nerve fibers is known from the anatomy of the rat and the vervet brain as described in the atlases, and was confirmed by the retardation images $\vert\sin\delta\vert$ shown below the transmittance images in \cref{fig:Rat_Vervet_Atlas}: regions with flat nerve fibers show larger retardation values than regions with steep nerve fibers. 
The mean transmittance values and the standard deviation for the evaluated green and yellow areas can be found in \cref{tab:Rat_Vervet_Transmittance}. 
\begin{table}[!htb]
	\caption{Mean transmittance values ($\overline{I_{\text{T,N}}}$) and standard deviation for the selected green and yellow areas in \cref{fig:Rat_Vervet_Atlas}. Areas belonging to the same structure were evaluated together (cg $\equiv$ cg1 $\cup$ cg2). 
		}
	\label{tab:Rat_Vervet_Transmittance} 
	\begin{center} 
		\begin{tabular}{c|cc|cc} 	\hline\hline
			\multicolumn{1}{c}{} 	& \multicolumn{2}{c}{Rat}					&  	\multicolumn{2}{c}{Vervet}		\\ 
				&	Coronal			&  Sagittal			& Coronal				& Sagittal							\\ \hline
			ac	&	--				& --				& -- 	& \fcolorbox{yellow}{white}{$0.11 \pm 0.04$} 		\\ 	
			cc	&	\fcolorbox{green}{white}{$0.33 \pm 0.16$}	& \fcolorbox{yellow}{white}{$0.09 \pm 0.04$}	
				& 	\fcolorbox{green}{white}{$0.24 \pm 0.09$} 	& \fcolorbox{yellow}{white}{$0.11 \pm 0.05$}		\\ 
			cg	&	\fcolorbox{yellow}{white}{0.15 $\pm$ 0.06}	& \fcolorbox{green}{white}{0.18 $\pm$ 0.09}
				&  	\fcolorbox{yellow}{white}{0.08 $\pm$ 0.02} 	& \fcolorbox{green}{white}{0.24 $\pm$ 0.09} 		\\ 
			df	&	\fcolorbox{yellow}{white}{$0.10 \pm 0.03$}	& -- & -- & --										\\ 
			f	&	--	& -- & \fcolorbox{yellow}{white}{$0.13 \pm 0.06$}& \fcolorbox{green}{white}{$0.26 \pm 0.10$}	\\ 
			fi	&	\fcolorbox{green}{white}{$0.25 \pm 0.10$}	& \fcolorbox{yellow}{white}{$0.15 \pm 0.07$} 	& -- & --	\\ 
			sm	&	\fcolorbox{yellow}{white}{$0.08 \pm 0.03$}	& -- & -- & --	\\[5pt]\hline\hline
		\end{tabular}
	\end{center}
\end{table}

In regions with flat nerve fibers, the mean transmittance values are larger ($\overline{I_{\text{T,N}}} \in [0.18, 0.33]$) than in regions with steep nerve fibers ($\overline{I_{\text{T,N}}} \in [0.08, 0.15]$). A region with flat (steep) nerve fibers which shows large (small) transmittance values in one section plane (coronal or sagittal), shows the opposite behavior in the corresponding orthogonal section plane. The difference is especially large when comparing the transmittance values of the \textit{corpus callosum} (a massive fiber tract connecting the two hemispheres of the brain) and the \textit{cingulum} (a C-shaped fiber structure running mostly perpendicular to the corpus callosum). In the coronal brain sections, the fibers of the cingulum (cg) run mostly perpendicular to the section plane and have about 55--67\% lower transmittance values than the fibers of the corpus callosum (cc) which lie mostly within the section plane. In the sagittal brain sections, the situation is exactly the opposite: the transmittance values of the corpus callosum are about 50--54\% less than the transmittance values of the cingulum. Fibers in the rat and vervet monkey brains show a similar pattern. 

As expected, images obtained from conventional bright-field transmission microscopy with unpolarized light show similar effects, not only in vervet but also in human brain sections (see \cref{fig:Human_PLI-vs-Zeiss} and Fig.\ S\ref{fig:Vervet_PLI-vs-Zeiss} in the Supplemental Material): regions with steep (out-of-plane) nerve fibers appear darker than regions with flat (in-plane) nerve fibers. 

\begin{figure}[!t]
	\centering
	\includegraphics[width=\columnwidth]{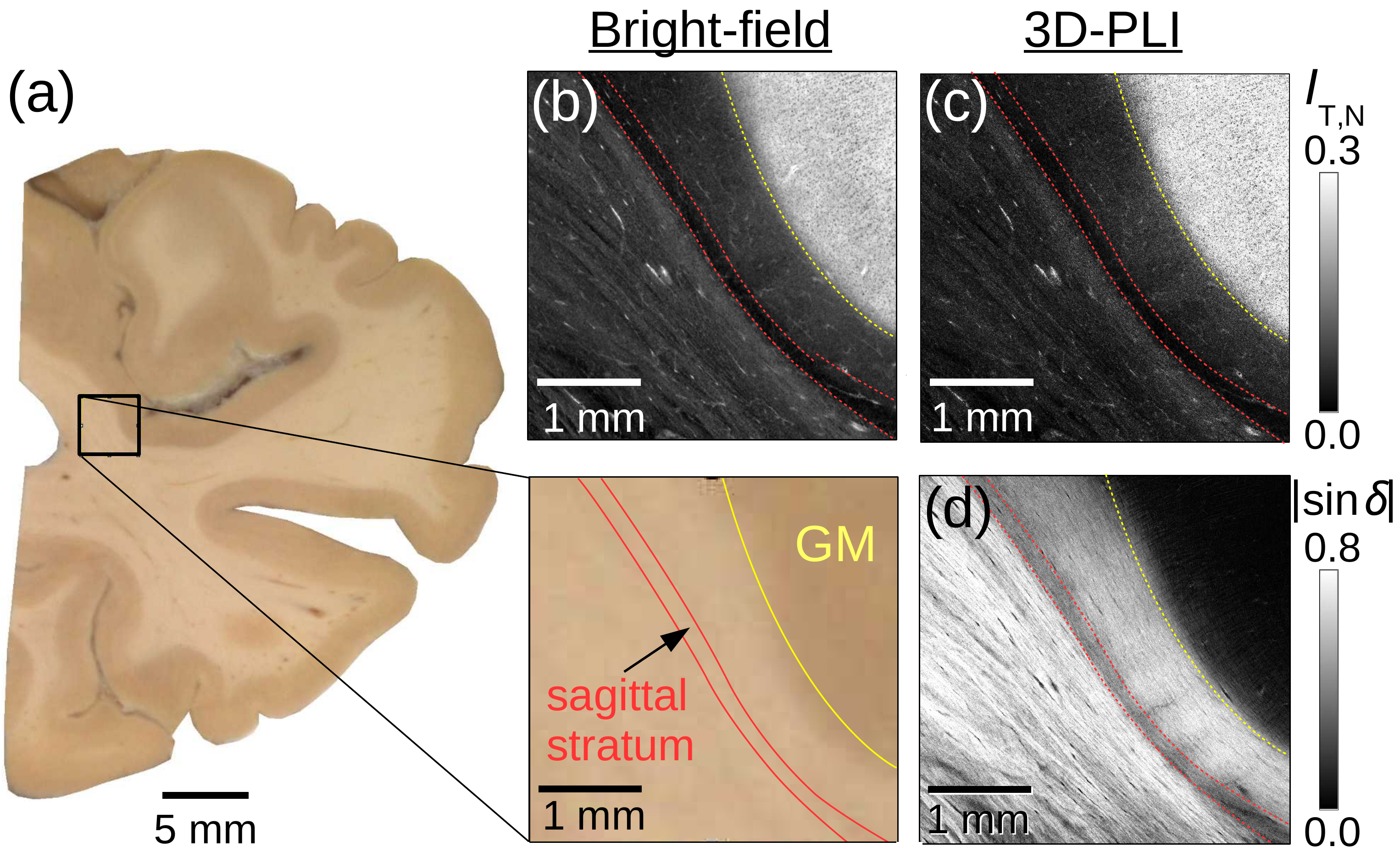}
	\caption{Bright-field transmission microscopy (see \cref{sec:bright-field}) vs.\ 3D-PLI measurement of a human brain section (right occipital lobe). (\textbf{a}) Photograph of the brain block surface before sectioning (blockface image). The enlarged region shows the \textit{sagittal stratum}, a white matter structure that runs mostly perpendicular to the section plane (GM = gray matter). (\textbf{b}) Bright-field transmission microscopy image of the same region with 0.91\,\um\ pixel size. (\textbf{c}) Normalized transmittance image of the same region obtained from a 3D-PLI measurement with 1.33\,\um\ pixel size. (\textbf{d}) Corresponding retardation image. Regions with steep out-of-plane fibers (\textit{sagittal stratum}) show lower transmittance (and retardation) values than the neighboring regions with non-steep fibers; the transmitted light intensity images obtained from bright-field transmission microscopy (b) and 3D-PLI (c) look similar.}
	\label{fig:Human_PLI-vs-Zeiss}
\end{figure}


\subsection{3D-reconstruction of transmittance images}
\label{sec:3D-transmittance}

\begin{figure}[!t]
	\centering
	\includegraphics[width=0.7\columnwidth]{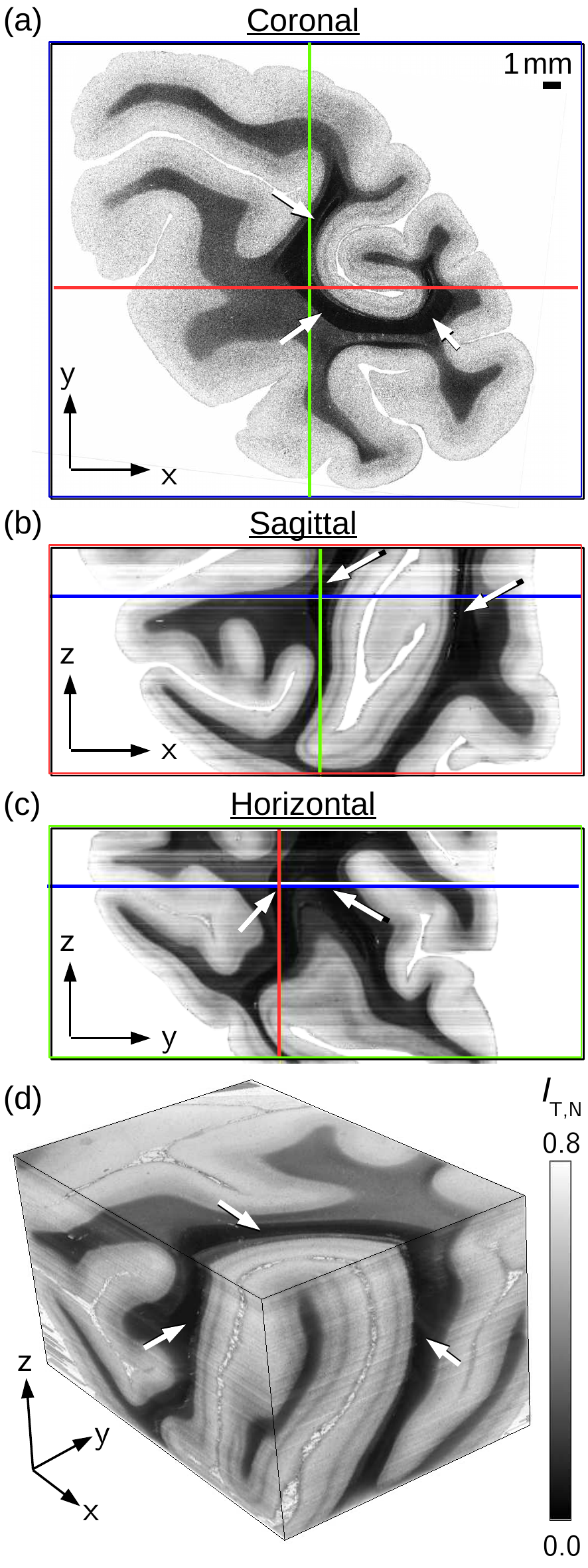}
	\caption{3D-reconstructed normalized transmittance images ($I_{\text{T,N}}$) of the right hemisphere of a vervet monkey brain (234 consecutive sections from the occipital lobe) obtained from 3D-PLI measurements with 1.33\,\um\ pixel size. The brain was cut along the coronal plane (xy-plane), the resulting brain sections were registered onto each other in the z-direction. (\textbf{a})-(\textbf{c}) Cross-sections of the 3D-volume shown along the coronal (xy), sagittal (xz), and horizontal (yz) plane. The colored lines indicate the position of the displayed xy-, xz-, and yz-planes. (\textbf{d}) Detail of the 3D-volume. The white arrows point to the \textit{sagittal stratum} -- a white matter structure that runs mostly perpendicular to the image plane (along the z-direction) and which appears much darker in the transmittance images than the surrounding tissue.}
	\label{fig:3D-Transmittance}
\end{figure}

So far, single brain sections from different specimens were compared to each other. To study the transmittance across several consecutive brain sections, the transmittance images of 234 coronal sections from the right hemisphere of a vervet monkey brain were registered onto each other using in-house developed software tools (see \cref{sec:3DPLI}).
\Cref{fig:3D-Transmittance} shows the 3D-reconstructed transmittance images along three orthogonal anatomical planes: coronal (a), sagittal (b), and horizontal (c), as well as a detail of the 3D-volume (d). The white arrows point to the \textit{sagittal stratum} -- a white matter structure with nerve fibers that are oriented mostly perpendicular to the image plane (along the z-direction), as can be seen in the sagittal and horizontal planes. The structure appears much darker in the transmittance images than the surrounding regions.
Thus, the observation that steep nerve fibers show lower transmittance values than flat nerve fibers is consistent across several consecutive brain sections.

\subsection{Transmittance contrast of nerve fiber bundles in mutually orthogonal planes}
\label{sec:transmittance-contrast}

\begin{figure*}[!t]
	\centering
	\includegraphics[width=1.5\columnwidth]{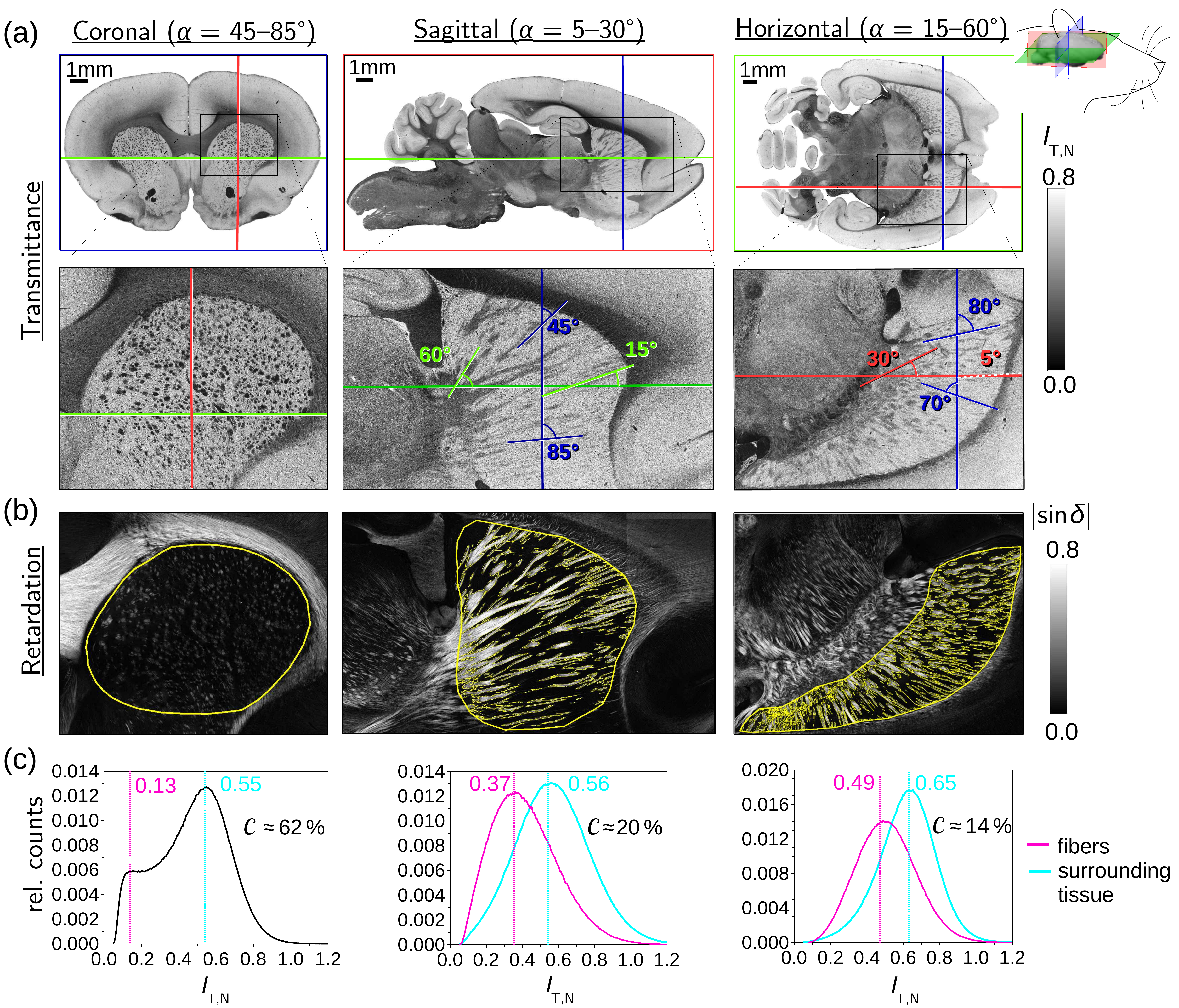}
	\caption{Transmittance contrast of nerve fiber bundles in mutually orthogonal anatomical planes. (\textbf{a}) Normalized transmittance images ($I_{\text{T,N}}$) of a coronal, sagittal, and horizontal rat brain section obtained from 3D-PLI measurements with 1.33\,\um\ pixel size. The colored lines indicate the approximate position of the section planes. The enlarged views show the region of the caudate putamen and the maximum and minimum angles under which the nerve fiber bundles are oriented with respect to the section planes. (\textbf{b}) Corresponding retardation images ($\vert\sin\delta\vert$) of the enlarged views. The image contrast was used to select regions with fibers and with surrounding tissue in the caudate putamen (yellow lines). (\textbf{c}) Histograms of the transmittance values ($I_{\text{T,N}}$) for the selected regions with nerve fibers (pink) and with surrounding tissue (cyan) in the caudate putamen. For the coronal brain section, the retardation image cannot be used to separate the fibers from the surrounding tissue because the fibers are oriented almost perpendicular to the image plane which leads to a small retardation signal and a small image contrast. Therefore, the histogram was computed over the whole selected region and the peak with lower (larger) transmittance was assumed to belong to fibers (surrounding tissue). The contrast values were computed from the respective peak values (numbers in pink and cyan) via: $\mathcal{C}$ = (max - min)/(max + min). The contrast for steep nerve fibers (coronal brain section) is much larger than for flat nerve fibers (sagittal and horizontal brain section).}
	\label{fig:Rat_CPu}
\end{figure*}

In the previous studies, we considered bulk tissue with densely packed nerve fibers where the fiber inclinations cannot be exactly determined. In regions with distinct fiber bundles, however, the inclination angles can be estimated by manually evaluating the course of the fiber bundles in different section planes. For this purpose, we selected a structure in the rat brain that contains several distinct nerve fiber bundles with different, well-defined inclination angles -- the so-called \textit{caudate putamen}. To estimate the inclination angles of the nerve fibers, we evaluated the course of the bundles in mutually orthogonal section planes (coronal, sagittal, horizontal), see \cref{fig:Rat_CPu}(a).
As the brain sections were obtained from different brains and might differ in tissue composition, the transmittance images cannot be directly compared to each other. To still enable a comparison between the transmittance of flat and steep nerve fibers, the transmittance values in regions with fibers were compared to the transmittance values in regions with surrounding tissue for each brain section, assuming that the transmittance of the surrounding tissue does not depend much on the choice of the section plane.

To separate the fiber bundles from the surrounding tissue, we used the image contrast of the retardation images (see yellow lines in \cref{fig:Rat_CPu}(b)). \Cref{fig:Rat_CPu}(c) shows the
corresponding histograms of the transmittance evaluated in regions with nerve fibers (pink) and in regions with surrounding tissue (cyan). As the coronal brain section contains mostly steep nerve fibers which yield low retardation values, the image contrast in the retardation image is not large enough to separate the fibers from the surrounding tissue. Therefore, we computed the histogram for the whole caudate putamen (area surrounded by yellow line) and assumed that the peak with lower (larger) transmittance belongs to nerve fibers (surrounding tissue).

From the minimum and maximum peak transmittance values of the histograms (pink and cyan numbers in \cref{fig:Rat_CPu}(c)), we computed the transmittance contrast $\mathcal{C} \equiv (I_{\text{T,max}} - I_{\text{T,min}})/(I_{\text{T,max}} + I_{\text{T,min}})$ between fiber bundles and surrounding tissue. For flat fiber bundles in the sagittal and horizontal brain sections, this contrast is much lower ($5^{\circ} \leq \alpha \leq 60^{\circ}$: $\mathcal{C} \approx 14$--$20\%$) than for steep fiber bundles in the coronal brain section ($45^{\circ} \leq \alpha \leq 85^{\circ}$: $\mathcal{C} \approx 62\%$). Assuming that the transmittance of the surrounding tissue is mostly independent of the fiber orientation, this demonstrates again that the transmittance values for steep nerve fibers are significantly lower than for flat nerve fibers.


\subsection{Transmittance vs. inclination of nerve fiber bundles in TPFM images}
\label{sec:transmittance-inclination-TPFM}

The previous studies were only qualitative and the observed differences in the transmittance might also be caused by a different tissue composition, \eg, a different density of nerve fibers. To quantitatively study how the transmittance of a brain section depends on the inclination angles of the enclosed nerve fibers, the exact underlying fiber structure of the investigated brain section needs to be known. Therefore, we measured the caudate putamen of a coronal mouse brain section both with 3D-PLI and with TPFM to identify the inclination angles of the enclosed fiber bundles (see inset in \cref{fig:Sim_Transmittance_vs_Inclination}(d) and Fig. S\ref{fig:PLI_vs_TPFM} in the Supplemental Material). 

The TPFM measurements were performed with a custom-made two-photon fluorescence microscope which achieves a resolution of $0.244 \times 0.244 \times 1\,\um^3$ and allows to perform an in-depth-scan of the brain section (see \cref{sec:TPFM} for more details). To obtain the inclination angles of the fiber bundles, the cross-sections of the bundles were determined in the first and the last slice of the TPFM image stack (cf.\ Fig.\ S\ref{fig:PLI_vs_TPFM}(d)). For each fiber bundle, the inclination angle was computed from the mid points of the corresponding cross-sections and from the thickness of the brain section (cf.\ Fig.\ S\ref{fig:PLI_vs_TPFM}(c)).
The fiber inclination and transmittance values were evaluated for 40 fiber bundles in the caudate putamen (see colored shapes in Fig.\ S\ref{fig:PLI_vs_TPFM}(b),(d)).

The scatter plot in \cref{fig:Sim_Transmittance_vs_Inclination}(d) shows the averaged transmittance values plotted against the determined fiber inclination angles.
Although the values are broadly distributed, the scatter plot shows a clear tendency towards a decrease in transmittance with increasing fiber inclination angle.
The values in orange belong to regions with lower fiber densities which might lead to overestimated transmittance values. However, the decrease in transmittance is also observed in regions with maximum fiber density (values in blue): while the mean transmittance values for flat nerve fibers ($\alpha < 50^{\circ}$) reach larger values ($ 0.1 < \overline{I_{\text{T,N}}} < 0.2$), the mean transmittance values for steep nerve fibers ($\alpha > 60^{\circ}$) are small ($\overline{I_{\text{T,N}}} < 0.05$).\\

All our experimental studies show that the transmittance of brain tissue decreases significantly (by more than 50\%) with increasing out-of-plane inclination angle of the enclosed nerve fibers.


\section{Simulation studies}
\label{sec:sim-studies}

Although the experimental studies clearly show that the transmittance depends on the inclination angle of the nerve fibers, they do not provide enough information to describe this effect in detail. Based on the experimental results alone, it is not possible to make any predictions or draw conclusions for the interpretation of measured data. To model and better understand the observed transmittance effects, we performed numerical simulations on artificial nerve fiber configurations. This has the advantage that the exact underlying fiber structure, and thus the inclination angles of the nerve fibers, are known -- also in bulk tissue with densely packed fibers.

As mentioned in \cref{sec:introduction}, the absorption coefficient of brain tissue is small so that the transmittance is expected to be mainly influenced by scattering. To study such complex light-matter interactions in microscopic detail, finite-difference time-domain (FDTD) algorithms are well suited. They discretize time and space, model the propagation of the light wave by approximating the spatial and temporal derivatives in Maxwell's curl equations by second-order central differences, and numerically compute the electromagnetic field components in space and time \cite{taflove,menzel2016,wilts2014,DeRaedt2012,wilts2012}.
As the mesh size of the spatial discretization needs to be much smaller than the wavelength (at most 25\,nm), simulations of tissue samples with dimensions of several micrometers are computationally very intense. We have developed a simulation framework that allows for the first time to use FDTD simulations to study larger samples of fibrous tissue. For this purpose, we used a simplified simulation model for the brain tissue samples and the optics of the imaging system.
We simulated the 3D-PLI measurement for various fiber configurations and evaluated the resulting transmittance values.

The artificial fiber configurations consist of about 700 fibers with uniformly distributed diameters between $1.0\,\um$ and $1.6\,\um$ and different fiber orientations. All fibers were generated in a volume of $30 \times 30 \times 30\,\um^3$ without intersections. The generation of the fiber configurations is described in \cref{sec:generation-fiber-config} in more detail.
Each fiber was represented by a simplified nerve fiber model, consisting of an inner axon with a constant radius and a surrounding myelin sheath with two layers, defined by different refractive indices (see \cref{sec:model-nerve-fibers} for motivation).

For the simulations of the 3D-PLI measurement, we used a conditionally stable FDTD algorithm to compute the propagation of the light wave through the tissue sample (artificial fiber configuration), described in more detail in \cref{sec:FDTD}. The resulting electric field components were processed with analytical methods taking all optical components of the polarimeter into account, including the objective lens (with numerical aperture NA $= 0.15$) and the camera detector (see \cref{sec:computation-light-intensities}).

The simulation studies were all performed for normally incident light with 550\,nm wavelength and for the simulation parameters listed in \cref{sec:sim-parameters}.
One simulation run (volume of $30 \times 30 \times 30\,\um^3$, mesh size of 25\,nm) took about 8000 core hours on the supercomputer JUQUEEN (using an MPI grid of $16 \times 16 \times 16$), allowing to perform many simulation runs with different parameters.
The accuracy of the simulation results is discussed in \cref{sec:error-estimation}.


\subsection{Simulated transmittance of inclined fibers}
\label{sec:simulated-transmittance}

\begin{figure*}[!t]
	\centering
	\includegraphics[width=1.5\columnwidth]{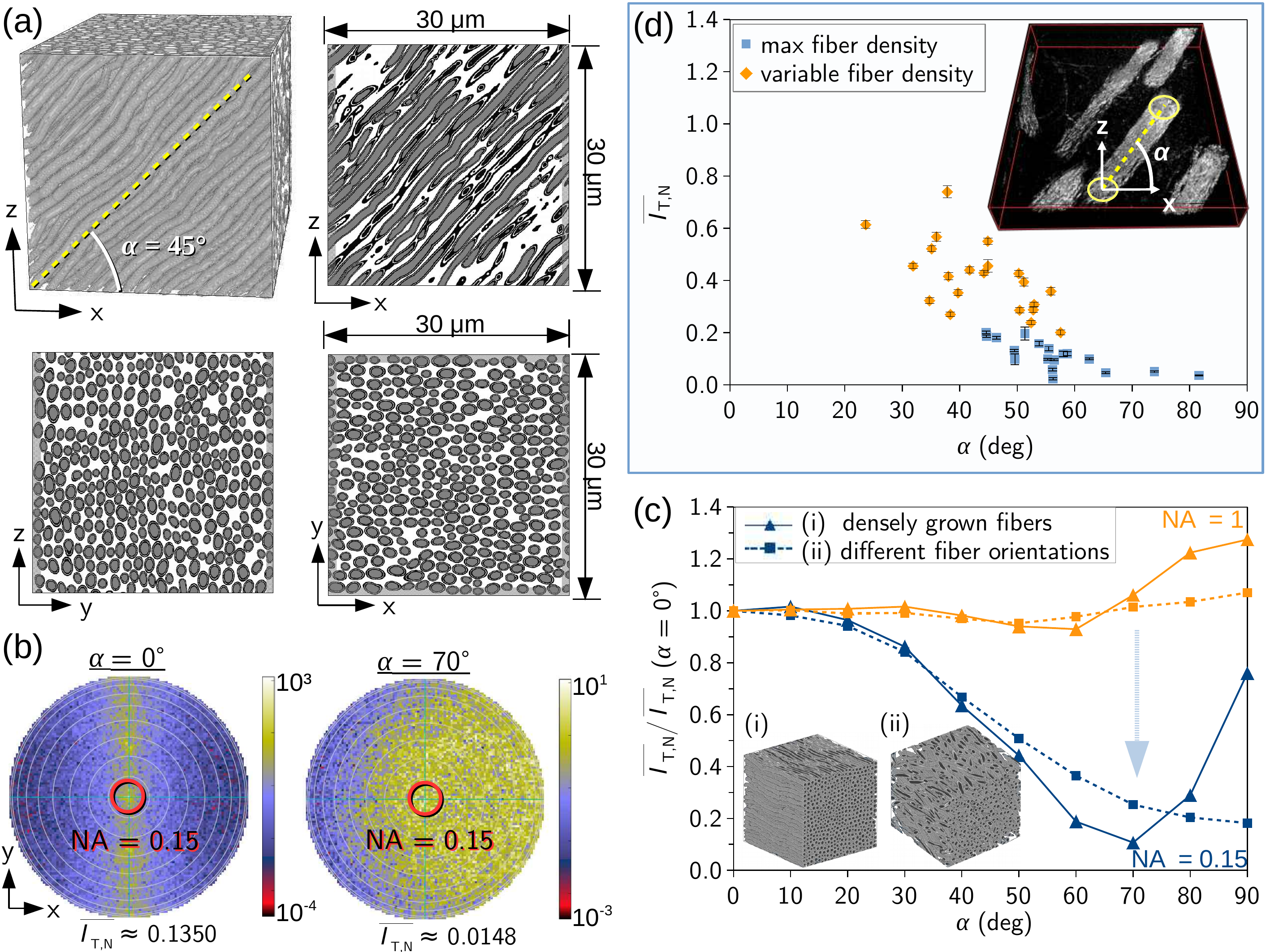}
	\caption{Transmittance of inclined fiber bundles. (\textbf{a}) 3D view and cross-sections through mid-planes for an artificial bundle of densely grown fibers shown exemplary for an inclination angle $\alpha=45^{\circ}$. (\textbf{b}) Light-scattering patterns obtained from 3D-PLI simulations for the bundle of densely grown fibers with $\alpha = 0^{\circ}$ and $70^{\circ}$. To compute the mean normalized transmittance values $\overline{I_{\text{T,N}}}$ for the numerical aperture of the imaging system (NA = 0.15), only wave vector angles $\theta_k \leq 8.6^{\circ}$ were considered (see red circles). The white circles represent steps of $\Delta\theta_k = 10^{\circ}$. (\textbf{c}) Simulated transmittance curves (mean transmittance $\overline{I_{\text{T,N}}}$ vs. inclination $\alpha$) for the bundle of densely grown fibers (i) and for a bundle with broad fiber orientation distribution (ii) for NA = 1 (orange curves) and NA = 0.15 (blue curves). The transmittance curves were normalized by the mean transmittance values of the horizontal bundles, respectively. The simulations were performed with the parameters specified in \cref{sec:sim-parameters}, using normally incident light with 550\,nm wavelength. (\textbf{d}) Mean normalized transmittance values $\overline{I_{\text{T,N}}}$ plotted against the nerve fiber inclination angles $\alpha$ determined respectively from 3D-PLI and TPFM measurements of nerve fiber bundles in a mouse brain section (see Fig.\ S\ref{fig:PLI_vs_TPFM} in the Supplemental Material). The values in blue belong to regions with similar (maximum) fiber density, the values in orange belong to regions with variable fiber density in which the transmittance might be overestimated. The error bars indicate the standard error of the mean for the evaluated transmittance values. Both experimental and simulated data show that the transmittance decreases with increasing fiber inclination.}
	\label{fig:Sim_Transmittance_vs_Inclination}
\end{figure*}

To better understand the inclination dependence of the transmittance that we observed in our experimental studies, we generated an artificial \textit{bundle of densely grown fibers} (see \cref{fig:Sim_Transmittance_vs_Inclination}(a) and \cref{sec:densely-grown}) for different inclination angles $\alpha = \{0^{\circ}, 10^{\circ}, \dots, 90^{\circ}\}$, and computed the transmittance from a simulated 3D-PLI measurement.

\Cref{fig:Sim_Transmittance_vs_Inclination}(b) shows the resulting scattering patterns (\ie, the intensity per wave vector angle $\theta_k$) of light transmitted through the sample for inclination angles $\alpha = 0^{\circ}$ and $70^{\circ}$. The white circles represent steps of $\Delta\theta_k = 10^{\circ}$, from $0^{\circ}$ (center) to $90^{\circ}$ (outer circle). The transmittance images and scattering patterns for all inclination angles can be found in Fig.\ S\ref{fig:Sim_TransmittanceImages} in the Supplemental Material.

For flat fibers ($\alpha < 45^{\circ}$), the light is mostly scattered under angles perpendicular to the principal axis of the fiber bundle (\ie, along the y-axis). For intermediate inclination angles, the light is scattered more and more in the direction of the fibers (\ie, in the positive x-direction). For an inclination angle of $70^{\circ}$, the light is broadly scattered in almost all directions (see \cref{fig:Sim_Transmittance_vs_Inclination}(b)). 

Due to the numerical aperture of the employed imaging system (NA $\approx$ 0.15), light scattered under angles larger than $\arcsin(\text{NA}) \approx 8.6^{\circ}$ does not contribute to the measured transmittance images.
To study the effect of the finite numerical aperture on the measured transmittance values, we simulated the imaging system without aperture (NA = 1) considering light scattered under all angles, and with aperture (NA = 0.15) considering only light scattered under angles $< 8.6^{\circ}$ (indicated by the red circles in \cref{fig:Sim_Transmittance_vs_Inclination}(b)).

\Cref{fig:Sim_Transmittance_vs_Inclination}(c) shows the resulting transmittance curves (mean values of the simulated transmittance images plotted against the inclination angles of the fiber bundle) for NA = 1 (orange curves) and NA = 0.15 (blue curves). The solid curves were obtained from the bundle of densely grown fibers (i) which has similar fiber orientations (the mode angle difference between the local fiber orientation vectors and the predominant orientation of the fiber bundle is less than $10^{\circ}$). The dashed curves were obtained for a \textit{bundle with broad fiber orientation distribution} (ii) which contains many different fiber orientations (the mode angle difference is about $25^{\circ}$, see \cref{sec:inhom-fiberbundle}). To enable a better comparison between horizontal fiber bundles ($\alpha = 0^{\circ}$) and vertical fiber bundles ($\alpha = 90^{\circ}$), all curves were divided by the mean transmittance value of the horizontal bundle, respectively. Figure S\ref{fig:Sim_TransmittanceCurves}(a),(c) in the Supplemental Material shows the (normalized) transmittance curves in separate figures.

For NA = 1, the transmittance for steep fibers is similar to or even slightly larger than the transmittance for flat fibers.
For NA = 0.15, the transmittance decreases significantly between $\alpha = 30^{\circ}$ and $\alpha = 70^{\circ}$. 
This implies that the observed decrease in transmittance is caused by the finite numerical aperture of the imaging system: for steep fibers, the light is scattered almost uniformly in all possible directions (cf.\ \cref{fig:Sim_Transmittance_vs_Inclination}(b) for $\alpha = 70^{\circ}$) so that the detected transmitted light intensity becomes minimal. In simulation studies with polarized light, we could show that the decrease in transmittance is independent of the direction of polarization (see \cref{fig:SimParameters_TransmittanceCurves}(a) in \cref{sec:error-estimation}) which suggests that the decrease is caused by isotropic (not by anisotropic) scattering of light.

While the transmittance for the bundle of densely grown fibers becomes minimal at $\alpha = 70^{\circ}$ for NA = 0.15 (the transmittance is 90\% less than the transmittance for the horizontal bundle) and the transmittance for vertical fibers ($\alpha = 90^{\circ}$) is only about 25\% less than for horizontal fibers ($\alpha = 0^{\circ}$), the transmittance for the bundle with broad fiber orientation distribution decreases monotonically with increasing fiber inclination angle and becomes minimal for vertical fibers (the transmittance for vertical fibers is more than 80\% less than for horizontal fibers).
Due to the broad fiber orientation distribution, the vertical bundle contains many fibers with inclinations between $60^{\circ}$ and $70^{\circ}$, which explains why the minimum transmittance is shifted to larger inclination angles.

Especially for the bundle with broad fiber orientation distribution, the simulated transmittance curves (\cref{fig:Sim_Transmittance_vs_Inclination}(c)) show a similar behavior as the measured transmittance values in the scatter plot (\cref{fig:Sim_Transmittance_vs_Inclination}(d)).\\

\begin{figure}[!t]
	\centering
	\includegraphics[width=0.6\columnwidth]{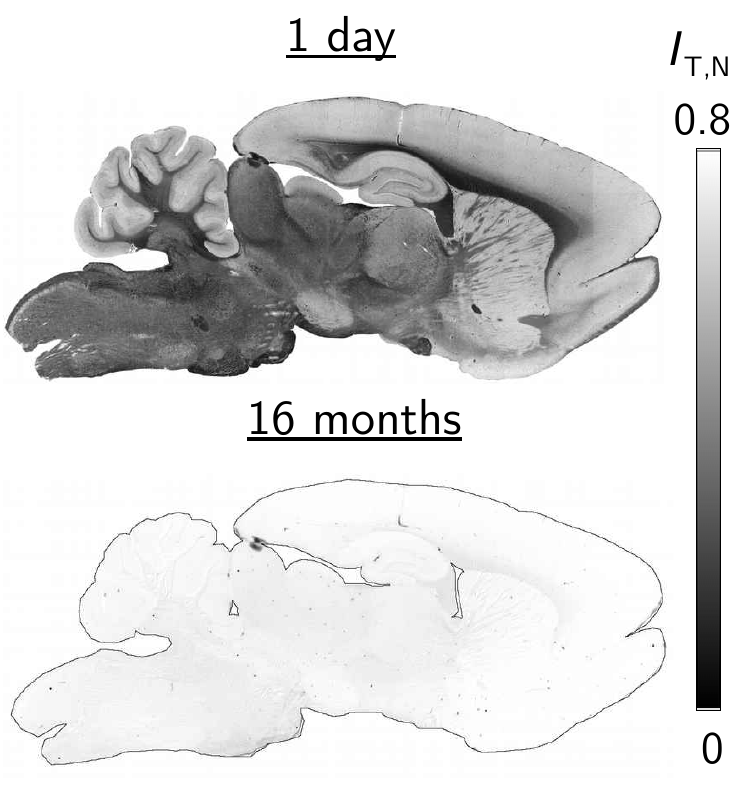}
	\caption{Normalized transmittance images $I_{\text{T,N}}$ of the sagittal rat brain section in \cref{fig:Rat_Vervet_Atlas}(a) obtained from a 3D-PLI measurement one day after tissue embedding and 16 months later. With increasing time after the tissue embedding, the brain section becomes more transparent.}
	\label{fig:Transmittance_vs_Time}
\end{figure}

As mentioned in the beginning, the brain sections used for the 3D-PLI measurements are embedded in glycerin solution.
With increasing time after this tissue embedding, we observed that the brain sections become more and more transparent, \ie, the transmittance increases (see \cref{fig:Transmittance_vs_Time}).
To enable optimal transmittance contrasts, the brain sections were therefore measured directly after the embedding and the simulations were performed assuming that the refractive indices of the nerve fibers correspond to given literature values (see \cref{sec:model-nerve-fibers}). 
With increasing time after the tissue embedding, the glycerin solution presumably soaks into the myelin sheaths which surround the axons. As the glycerin solution has a lower refractive index than the myelin lipids, the effective refractive index of the myelin sheaths is therefore expected to decrease. Figure S\ref{fig:Sim_TransmittanceCurves}(b) in the Supplemental Material shows the resulting transmittance curves for the bundle of densely grown fibers with a reduced myelin refractive index: the mean transmittance for steep fibers ($\alpha=70$--$80^{\circ}$) is only about 30\% less than the mean transmittance for horizontal fibers, and the absolute transmittance values become larger. This corresponds to the experimental observation that the transmittance increases with increasing time after the tissue embedding.

\subsection{Simulated transmittance of crossing fibers}
\label{sec:sim-transmittance-crossing}

Our previous simulation studies have shown that the transmittance strongly depends on the out-of-plane fiber inclination angle. Hence, the transmittance could be used to distinguish out-of-plane fibers from in-plane crossing fibers, which both yield small birefringence signals and can to date not be distinguished in standard 3D-PLI measurements.
To study this in more detail, we simulated the transmittance of horizontal (in-plane) crossing fibers for different crossing angles and compared the results to the transmittance of steep (out-of-plane) fibers.

\begin{figure*}[!htb]
	\centering
	\includegraphics[width=1.5\columnwidth]{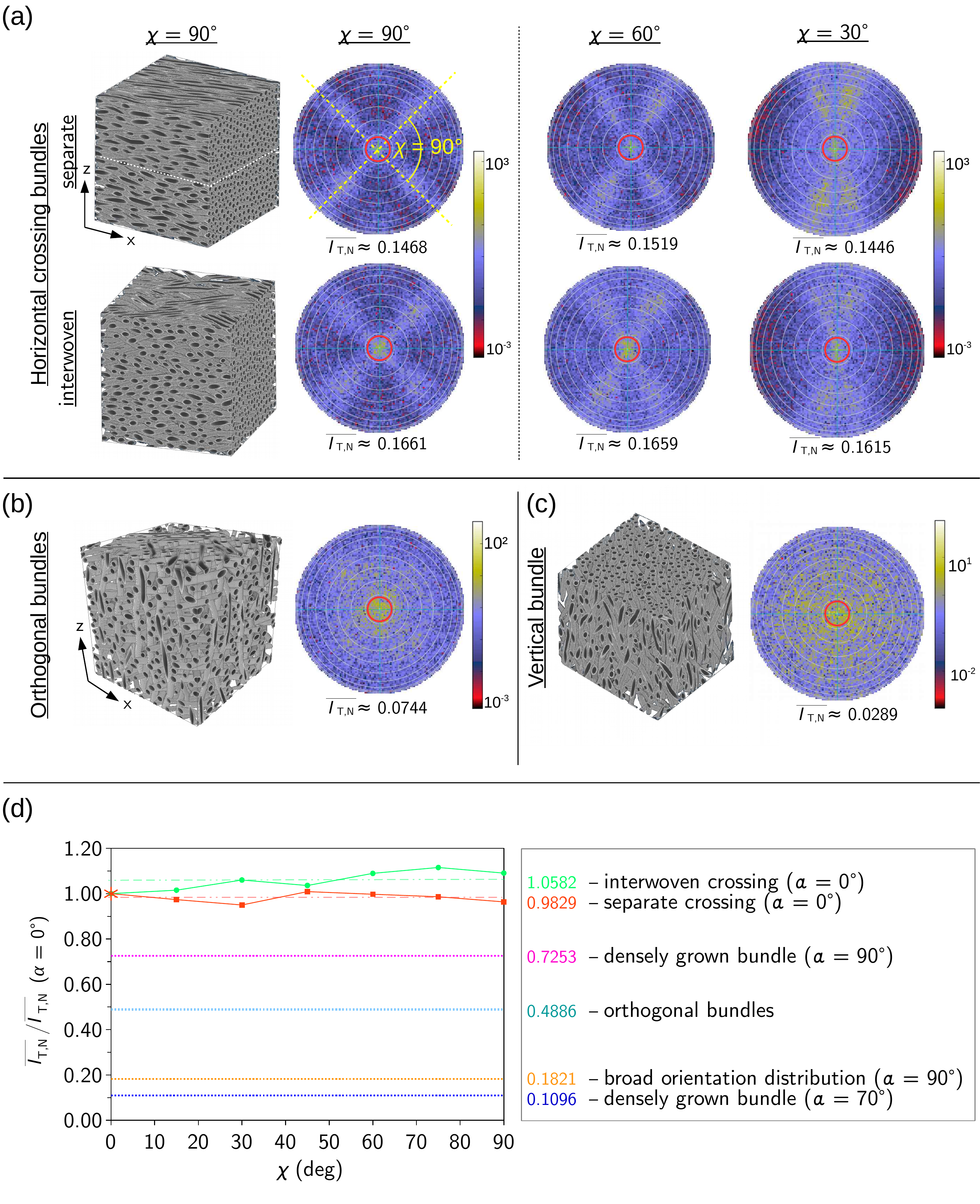}
	\caption{Simulated transmittance of crossing fibers. (\textbf{a})-(\textbf{c}) 3D views and light-scattering patterns for (a) horizontal crossing fibers (separate bundles in the upper row, interwoven bundles in the lower row) with crossing angle $\chi$, (b) three mutually orthogonal, interwoven fiber bundles, and (c) a vertical fiber bundle with broad fiber orientation distribution (cf.\ \cref{fig:Sim_Transmittance_vs_Inclination}(c)(ii)). The mean normalized transmittance values $\overline{I_{\text{T,N}}}$ were computed from a simulated 3D-PLI measurement (with numerical aperture NA = 0.15). The simulations were performed with the parameters specified in \cref{sec:sim-parameters}, using normally incident light with 550\,nm wavelength. (\textbf{d}) Mean transmittance values for different crossing angles $\chi$ shown for in-plane crossing (solid curves) and out-of-plane fiber configurations (densely dotted curves). For better comparison, the values were divided by the mean transmittance value of the corresponding horizontal fiber bundle for $\chi=0^{\circ}$. Apart from the fiber configurations shown in this figure, the mean transmittance values are also displayed for the bundle of densely grown fibers (see \cref{fig:Sim_Transmittance_vs_Inclination}(c)(i)) for $\alpha = 70^{\circ}$ and $90^{\circ}$. While the scattering patterns show the crossing angles of in-plane crossing fibers (see (a)), the mean transmittance is mostly independent of the crossing angle and larger than the mean transmittance of out-of-plane fibers (see (d)).}
	\label{fig:Sim_CrossingFibers}
\end{figure*}

The horizontal crossing fibers were generated as separate and interwoven fiber bundles (see \cref{sec:inhom-fiberbundle} and \cref{fig:CrossingFibers}(a)-(b)) with different crossing angles $\chi = \{0^{\circ}, 15^{\circ}, \dots, 90^{\circ} \}$.
\Cref{fig:Sim_CrossingFibers}(a) shows the resulting scattering patterns and normalized mean transmittance values $\overline{I_{\text{T,N}}}$ for NA = 0.15 and $\chi = \{90^{\circ}, 60^{\circ}, 30^{\circ} \}$. 

The scattering patterns of separate and interwoven crossing fiber bundles look similar for all crossing angles. The underlying fiber configuration, \ie, the crossing angle of the fiber bundles, is clearly visible in all scattering patterns. 
The mean transmittance values for the interwoven fiber bundles are up to 13\% larger than those for the separate fiber bundles and in both cases mostly independent of the crossing angle.

\Cref{fig:Sim_CrossingFibers}(b) shows the scattering pattern for three mutually orthogonal, interwoven fiber bundles. The fiber configuration is similar to the horizontal $90^{\circ}$-crossing, interwoven fiber bundles, but one third of the fibers is oriented in the z-direction (see \cref{sec:inhom-fiberbundle} and \cref{fig:CrossingFibers}(c)). 
This configuration leads to lower transmittance values than the horizontal crossing fibers.

\Cref{fig:Sim_CrossingFibers}(c) shows the scattering pattern for a vertical fiber bundle (bundle with broad fiber orientation distribution and $\alpha = 90^{\circ}$, cf.\ \cref{fig:Sim_Transmittance_vs_Inclination}(c)(ii)).
The scattering pattern looks clearly different from the scattering pattern of the horizontal crossing fibers and the mean transmittance is much lower.

\Cref{fig:Sim_CrossingFibers}(d) shows the mean transmittance values of the different fiber bundles for NA = 0.15 plotted against the crossing angle $\chi$ (in the case of horizontal crossing fibers). For better comparison, the values were divided by the mean transmittance value of the corresponding horizontal fiber bundle (for $\chi=0^{\circ}$), respectively.
The solid curves belong to the horizontal crossing fibers (separate and interwoven bundles), the densely dotted lines below belong to fiber constellations that contain vertical or steep fibers: the bundle of densely grown fibers for $\alpha = 90^{\circ}$ and $70^{\circ}$, the mutually orthogonal fiber bundles, and the bundle with broad fiber orientation distribution for $\alpha = 90^{\circ}$.\\

The transmittance curves of horizontal crossing fibers are similar for separate and interwoven fiber bundles. While the mean transmittance of the separate crossing fibers corresponds more or less to the mean transmittance of the horizontal fiber bundle for $\chi = 0^{\circ}$, the transmittance values of the interwoven crossing fibers slightly increase with increasing crossing angle (by max.\ 11\%). 

For all simulated fiber bundles that contain vertical or steep fibers, the mean transmittance values (densely dotted lines) are more than 26\% less than for the horizontal crossing fibers (solid lines).
For interwoven crossing fibers, the transmittance value is reduced by more than one half when the horizontal crossing fibers are combined with a vertical fiber bundle (orthogonal bundles).
For the vertical bundle with broad fiber orientation distribution and the steep bundle of densely grown fibers (with $\alpha = 70^{\circ}$), the difference between the transmittance values is especially large: the transmittance is about 80--90\% less than for the horizontal crossing fibers.\\

Our simulations of crossing fiber bundles have shown that the transmittance for horizontal fibers is mostly independent of the crossing angle between the bundles and much larger than the transmittance for vertical fibers. This suggests that the transmittance values can be used to distinguish between horizontal crossing and vertical fibers in 3D-PLI measurements, and to detect vertical fibers within fiber crossings.


\section{Combination of experimental and simulation studies}
\label{sec:exp-sim}

\begin{figure*}[!t]
	\centering
	\includegraphics[width=1.5\columnwidth]{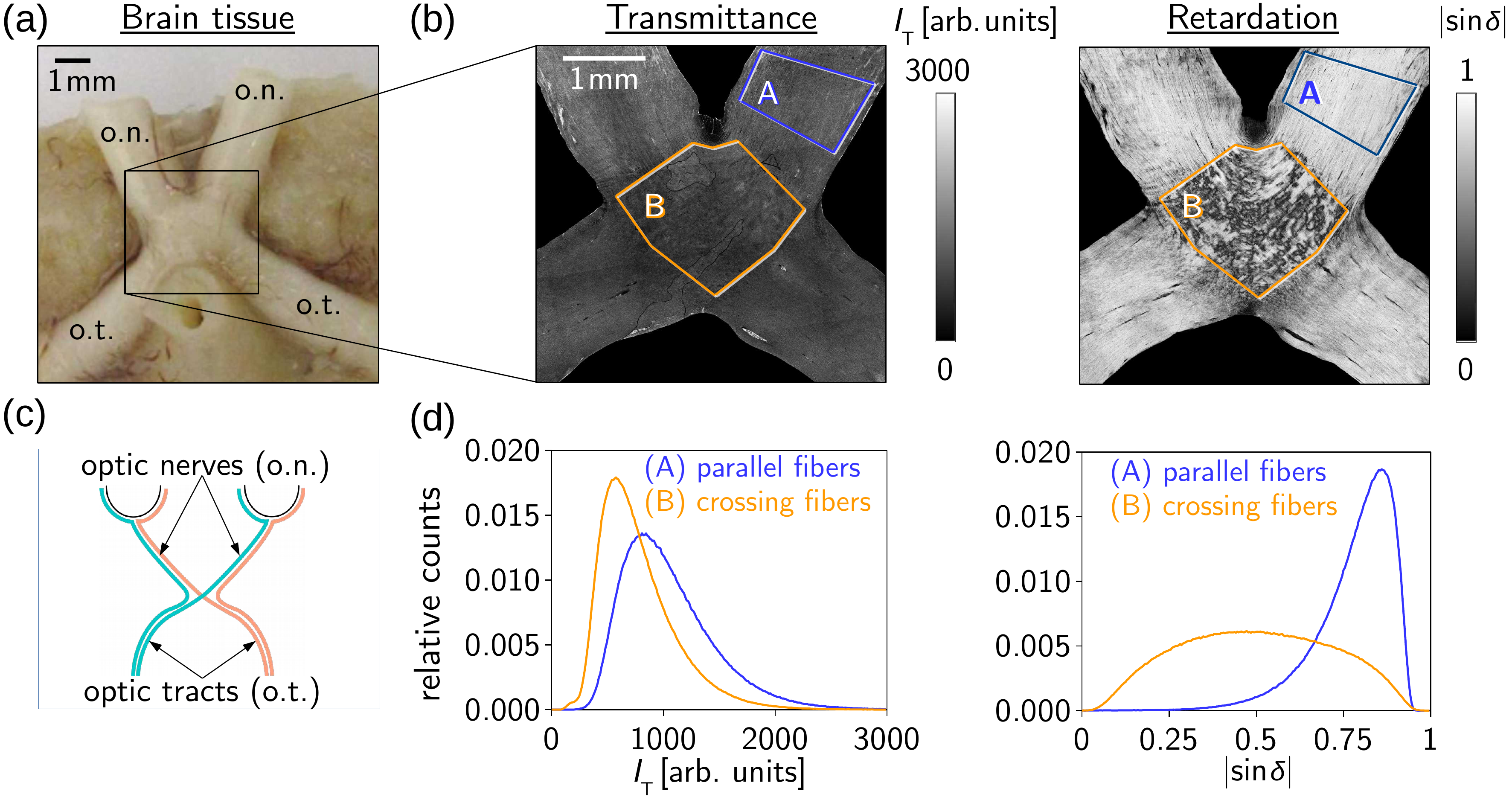}
	\caption{Crossing nerve fibers in the optic chiasm of a hooded seal: (\textbf{a}) brain tissue before sectioning, (\textbf{b}) unnormalized transmittance and retardation images of the middle brain section obtained from 3D-PLI measurements with 1.33\,\um\ pixel size, (\textbf{c}) schematic drawing of the optic chiasm consisting of optic tracts (o.t.) and optic nerves (o.n.), (\textbf{d}) normalized histograms of the transmittance image ($I_{\text{T}}$) and retardation image ($\vert\sin\delta\vert$) for a region with mostly parallel fibers (blue) and a region with nearly $90^{\circ}$-crossing fibers (orange). Unlike the retardation, the transmittance does not depend on the crossing angles of the nerve fibers, only on the tissue density. More information about the sample can be found in \textsc{Dohmen} \ea\ \cite{dohmen2015} (Figs.\ (a) and (c) were adapted from Figs.\ 1B and 5B0 in \cite{dohmen2015} Copyright (2015), with permission from Elsevier).}
	\label{fig:OpticChiasm}
\end{figure*}

In this section, we combine the results from the experimental studies and the simulation studies to develop a classification for brain regions with low birefringence signals that cannot be distinguished by 3D-PLI measurement. The simulations in \cref{sec:sim-transmittance-crossing} have shown that the transmittance does not depend on the crossing angle between in-plane nerve fibers. First, we verify this prediction by investigating the optic chiasm of a hooded seal \cite{dohmen2015} -- a region that contains fibers with crossing angles of around $90^{\circ}$ in the image plane (see \cref{fig:OpticChiasm}). 

While the retardation values in the region with crossing fibers (region B in orange) are broadly distributed (the birefringence signals of crossing fibers cancel out), the transmittance values in this region show a similar distribution as in a region with mostly parallel fibers (region A in blue), see histograms in \cref{fig:OpticChiasm}(d). 

The peak transmittance value of region B is slightly lower than in region A because the number of fibers in the crossing region (two crossing bundles) is larger than in the region with parallel fibers (one bundle). Thus, the transmittance depends on the tissue density, but not on the crossing angles between the nerve fibers -- as predicted by the simulations in \cref{sec:sim-transmittance-crossing}.

\begin{figure*}[!t]
	\centering
	\includegraphics[width=1.45\columnwidth]{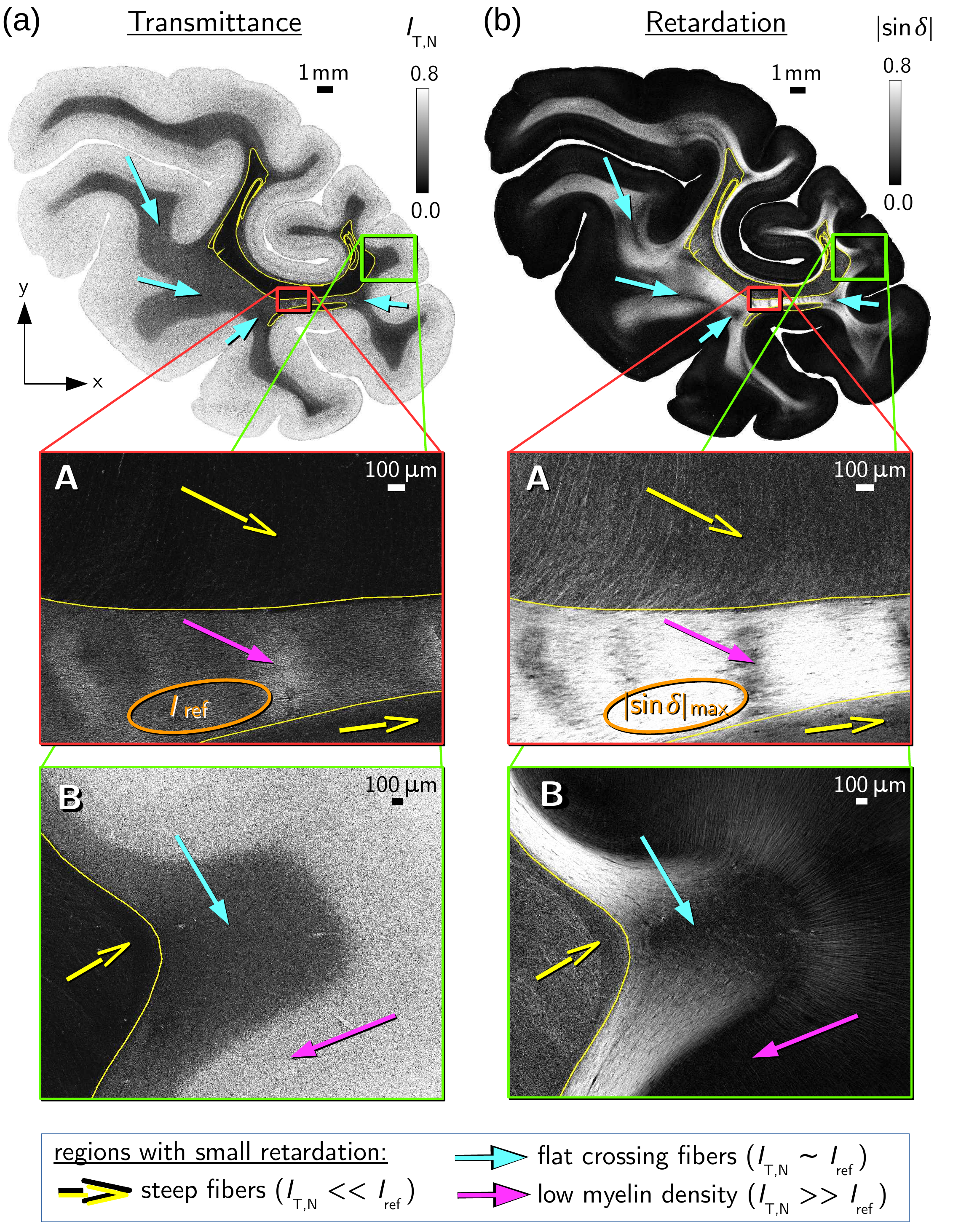}
	\caption{Combined analysis of transmittance and retardation images allowing to distinguish between brain regions with in-plane crossing and out-of-plane nerve fibers. The figure shows the normalized transmittance image ($I_{\text{T,N}}$) and the retardation image ($\vert\sin\delta\vert$) of a coronal section through the right hemisphere (occipital lobe) of a vervet brain (cf.\ \cref{fig:3D-Transmittance}(a)). The transmittance in the region with maximum retardation (orange ellipse) is used as a threshold value ($I_{\text{ref}} \equiv I_{\text{T,N}}(\vert\sin\delta\vert_{\text{max}})$). Regions with small retardation values and notably lower transmittance values ($I_{\text{T,N}} \ll I_{\text{ref}}$, yellow arrows and regions surrounded by a yellow line) are expected to contain steep (out-of-plane) fibers. Regions with small retardation values and similar transmittance values ($I_{\text{T,N}} \sim I_{\text{ref}}$, cyan arrows) are expected to contain flat (in-plane) crossing fibers. Regions with small retardation values and larger transmittance values ($I_{\text{T,N}} \gg I_{\text{ref}}$, magenta arrows) belong to regions with low fiber density, \ie, regions with a large amount of unmyelinated axons or surrounding tissue.}
	\label{fig:Transmittance_vs_Retardation}
\end{figure*}

To demonstrate that the transmittance can be used to classify regions with small birefringence signals (\ie, small retardation values obtained from 3D-PLI measurements) into regions with in-plane crossing fibers, regions with steep fibers, and regions with low fiber density, the predictions obtained from the simulation studies in \cref{sec:sim-studies} were applied to experimental data (see \cref{fig:Transmittance_vs_Retardation}).

As the transmittance depends on absorption, the region with maximum absorption was determined as a reference:
The retardance $\delta$ of brain tissue increases with decreasing fiber inclination angle $\alpha$ and with increasing thickness $d$ of birefringent tissue components ($\delta \propto d\,\dn \, \cos^2\alpha$, where $\dn$ is the birefringence of the tissue \cite{MAxer2011_1}). 
Assuming that a brain section contains all possible nerve fiber configurations, the region with maximum retardation signal $|\sin\delta|_{\text{max}}$ (orange ellipse in \cref{fig:Transmittance_vs_Retardation}A) is therefore expected to contain mostly horizontal parallel fibers ($\alpha \approx 0^{\circ}$) with a high fiber density (max.\ $d\,\dn$) and thus to cause a maximum of absorption. Regions with even lower transmittance values are accordingly expected to contain steep (out-of-plane) fibers which increase the scattering and thus the attenuation of light.

By comparing the normalized transmittance values ($I_{\text{T,N}}$) of regions with small retardation values to the transmittance of the region with maximum retardation ($I_{\text{ref}} \equiv I_{\text{T,N}}(|\sin\delta|_{\text{max}}))$, the regions can be classified into three categories (see \cref{fig:Transmittance_vs_Retardation}):
\begin{enumerate}
	\item $I_{\text{T,N}} \ll I_{\text{ref}}$\,: regions with notably lower transmittance values are expected to contain steep (out-of-plane) fibers (see yellow arrows and regions surrounded by a yellow line),
	\item $I_{\text{T,N}} \sim I_{\text{ref}}$\,: regions with similar transmittance values are expected to contain flat (in-plane) crossing fibers (see cyan arrows),
	\item $I_{\text{T,N}} \gg I_{\text{ref}}$\,: regions with notably larger transmittance values are expected to have a lower fiber density (see magenta arrows).
\end{enumerate}
For regions with slightly lower or larger transmittance values, an unambiguous classification is not possible.
Provided that the region with maximum retardation has the largest tissue absorption, lower transmittance values can only be caused by steep fibers. Similar transmittance values, however, could also be caused by a small number of steep fibers, and larger transmittance values could be caused by a small number of in-plane crossing fibers (or a smaller number of steep fibers). A classification by means of retardation and transmittance values can therefore only serve as an indication of the underlying fiber configuration and should always be considered in addition to individual tissue characteristics.
As the transmittance depends on the tissue preparation, the combined analysis of transmittance and retardation should only be performed section-wise.
Brain atlases and 3D-reconstructed images (cf.\ \cref{fig:3D-Transmittance}) validate the classification of regions in  \cref{fig:Transmittance_vs_Retardation}.

\section{Discussion and Conclusion}
\label{sec:discussion}

Conventional bright-field transmission microscopy measurements of fibrous tissue samples provide usually only 2D information about the underlying fiber architecture. This considerably limits the application of these methods in the analysis of three-dimensional fiber structures, \eg, when studying the complex architecture of nerve fibers in the brain.
In this paper, we show both in experimental and simulation studies that the scattering of light enables a more enhanced interpretation of transmission microscopy images, providing additional structural information about the three-dimensional fiber architecture in brain tissue samples.

First, we exploited various techniques to derive three-dimensional structural information from brain tissue samples with unknown substructures.
Our experimental studies on brain sections from different species (rodent, monkey, and human, see \cref{sec:exp-studies}) have shown that the polarization-independent transmitted light intensity (transmittance) significantly decreases with increasing out-of-plane inclination angle of the enclosed nerve fibers (by more than 50\%). Using finite-difference time-domain (FDTD) simulations, we could successfully model this effect and show that the decrease in transmittance is caused by polarization-independent (isotropic) light scattering and by the finite numerical aperture of the imaging system (see \cref{sec:simulated-transmittance}).
Polarization-dependent light scattering which leads to diattenuation (polarization-dependent attenuation of light) cannot explain the observed transmittance effect because the diattenuation of brain tissue was shown to be small \cite{menzel2017,menzel2019}.

Furthermore, we could use the FDTD simulations to explain the increasing transparency of brain tissue samples with increasing time after tissue embedding (see \cref{fig:Transmittance_vs_Time}): when the embedding solution soaks into the surrounding myelin sheaths of the nerve fibers, this leads to an equalization of the effective refractive indices and thus to reduced scattering, which increases the transparency of the tissue.

Note that both measured and simulated transmittance values depend on many parameters such as the homogeneity of brain tissue or the density of nerve fibers, which are not easily accessible. As long as the exact underlying tissue structure is not known, our simulation results can therefore only serve as a qualitative prediction for the interpretation of measured data.

As the simulated samples are only characterized by their geometry and refractive indices, biological and non-biological samples with comparable fibrous structures (\eg, muscle fibers, collagen, artificial fibers) are expected to show similar transmittance effects. To increase the transmittance contrast between flat and steep fiber structures, the embedding solution should have a different refractive index than the fibers (cf.\ Fig.\ S\ref{fig:Sim_TransmittanceCurves}(a)-(b) in the Supplemental Material). 

The observed transmittance effects are mostly independent of the polarization so that standard transmission microscopy techniques can be used to obtain 3D information about the underlying fiber configurations, without need to change the experimental setup or to repeat measurements. This greatly enhances the application of conventional bright-field transmission microscopy which is available in many laboratories and so far only used to gain 2D structural information.

When generating a detailed model of the nerve fiber architecture in the brain, the reconstruction of brain regions with crossing nerve fibers poses a major challenge.
With 3D-Polarized Light Imaging (3D-PLI), it is generally not possible to distinguish brain regions with in-plane crossing fibers from regions with out-of-plane fibers or from regions with low fiber densities, because they all yield low birefringence signals. 
Using FDTD simulations, we could show that the transmittance of in-plane fiber configurations does not depend on the crossing angle between the fibers (see \cref{sec:sim-transmittance-crossing}).
Applying the predictions of our simulation studies to experimental data, we could demonstrate that a combined analysis of transmittance and retardation (strength of the birefringence signal) enables to distinguish between these regions (see \cref{sec:exp-sim}). 
Our simulations also revealed that the transmittance can be used to detect out-of-plane fibers in regions with in-plane crossing fibers, which is not possible with current techniques.
The combined analysis of transmittance and retardation images can also be applied to past 3D-PLI measurements in order to validate and -- if necessary -- correct the reconstructed fiber orientations.

Apart from the fiber inclination, the transmitted light intensity reveals much more information about the underlying tissue structure when studying the exact pattern of the scattered light. Our simulation studies in \cref{sec:sim-transmittance-crossing} have shown that the scattering pattern can be used, for example, to identify the crossing angle of nerve fiber bundles, which is not easily accessible with current measurements. How the scattering pattern is related to the exact underlying fiber structure and tissue homogeneity will be addressed in future studies.
Major features of the scattering pattern (like the fiber crossing angle) can be determined by simply placing an aperture between light source and sample and measuring the transmitted light intensity for different positions of the aperture.

The FDTD simulations proved to be a valuable and reliable tool in many aspects: they allow to better understand the interaction of polarized light with brain tissue, to find explanations for the observed transmittance effects, to make general predictions, and to improve the measurement procedure and analysis. 
In contrast to previous top-down simulation approaches of 3D-PLI that model the optical properties of the nerve fibers by series of Jones matrices (\textit{simPLI} \cite{dohmen2015,menzel2015}), the FDTD simulations solve Maxwell's equations and allow to model more complex effects like the scattering of light, but they require much more computing time.

Most nerve fibers in the brain are surrounded by a so-called myelin sheath, which consists of multiple layers with 3--5\,nm thickness (see \cref{sec:model-nerve-fibers}). If the exact layered structure of the myelin sheath is modeled, the mesh size in the simulations can be at most 3\,nm. In this case, the simulation of a single nerve fiber with 1\,\um\ diameter consumes almost 290\,000 core hours (see \cref{sec:diff-myelinlayers}). To enable the simulation of larger tissue samples with various simulation parameters, we developed a simplified nerve fiber model with double myelin layers and a simplified model for the imaging system (the incoherent and diffusive light source was modeled by monochromatic light with normal incidence). We could show that these simplified models still reproduce the observed transmittance effects and that our results are not sensitive to small changes in the simulation parameters (see \cref{sec:error-estimation} and \cite{menzel}), so our model is a good compromise between accuracy and computing time.
The developed simulation framework can easily be adapted to microscopy techniques with different optics (numerical aperture, wavelength, polarization, etc.) and to other species and tissue types. 
Further brain tissue components like glial cells can easily be added to the simulation model \cite{ginsburger2019}.

In summary, we have developed and successfully applied a versatile simulation framework for transmission microscopy measurements of fibrous tissue samples that allows to study light scattering in larger samples like brain tissue, using finite-difference time-domain simulations.
We have demonstrated both in experimental and simulation studies on various brain tissue samples that the polarization-independent transmitted light intensity (transmittance) provides information about the 3D orientation of the enclosed fibers, allowing to use simple bright-field transmission microscopy to study three-dimensional fiber structures. Finally, we could show that the transmittance can be used to classify brain regions with low birefringence signals without changing the experimental setup or repeating measurements. This enables a more enhanced interpretation of three-dimensional nerve fiber architectures in the brain.


\begin{acknowledgments}
	We thank Markus Cremer, Christian Rademacher, and Patrick Nysten for the preparation of the histological brain sections, David Gr\"{a}{\ss}el and Isabelle Mafoppa Fomat for the 3D-PLI measurements, Philipp Schl\"{o}mer for generating the transmittance and retardation images, Martin Schober, Marcel Huysegoms, and Sascha M\"{u}nzing for image registration, Felix Matuschke for developing the algorithms to generate the fiber configurations, Sebastian Bludau for the bright-field transmission microscopy measurements, Andreas Wree for providing the human brain sample, and Karl Zilles and Roger Woods for collaboration in the vervet brain project.
	This work has received funding from the Helmholtz Association portfolio theme \textit{Supercomputing and Modelling for the Human Brain}, from the European Union's Horizon 2020 Research and Innovation Programme under Grant Agreement No.\ 7202070 and 785907 (\textit{Human Brain Project} SGA1 and SGA2), and from the National Institutes of Health under grant agreement No.\ R01MH092311 and 5P40OD010965.
	We gratefully acknowledge the computing time granted through JARA-HPC on the supercomputer \textit{JURECA} \cite{jureca} and \textit{JUQUEEN} \cite{juqueen} at Forschungszentrum J\"{u}lich.
	M.M. designed the study with help from M.A. and K.M., analyzed the measurements and carried out the simulations. M.M., M.A., H.D.R., and K.M. contributed to the interpretation of the data and provided theoretical considerations. H.D.R. and K.M. provided the FDTD software. I.C., L.S., and F.S.P.\ produced the TPFM measurements.
	K.A.\ contributed to the anatomical content of the study. M.M. wrote the paper with revisions from M.A., H.D.R., K.A., and K.M.	
\end{acknowledgments}


\appendix


\section{Measurement methods}
\label{sec:meas-methods}

\subsection{Preparation of brain sections}
\label{sec:brain-preparation}
The experimental studies in \cref{sec:exp-studies,sec:exp-sim} were performed on sections from a human brain (male, 87 years old), as well as on brain sections from vervet monkeys (African green monkey: \textit{Chlorocebus aethiops sabaeus}, male, between one and two years old), rats (\textit{Wistar}, male, three months old), mice (\textit{C57BL/6}, male, six months old), and a hooded seal \cite{dohmen2015}. 
All animal procedures were approved by the institutional animal welfare committee at Forschungszentrum J\"{u}lich GmbH, Germany, and are in accordance with European Union (National Institutes of Health) guidelines for the use and care of laboratory animals.
The human brain was acquired in accordance with the local ethic committee of the University of Rostock, Germany. A written informed consent of the subject is available.

The brains were removed from the skull within 24 hours after death, immersed in a buffered solution of 4\% formaldehyde for several weeks, immersed for several days in solutions of 10\% and 20\% glycerin combined with 2\% Dimethyl sulfoxide for cryoprotection, dipped in cooled isopentane for several minutes, and deeply frozen. The frozen brains were cut with a cryostat microtome (\textit{Leica Microsystems}, Germany) at a temperature of -30\,$^{\circ}$C into sections of 60\,\um. The brain sections were mounted on cooled glass slides, embedded in 20\% glycerin solution, covered by a cover glass, sealed with lacquer, and weighted for several hours to prevent the development of air bubbles. The sections were measured one day after embedding to obtain optimal transmittance images.
\vspace{-0.3cm}


\subsection{3D-PLI measurement}
\label{sec:3DPLI}

The 3D-PLI measurements were performed with a high-resolution \textit{Polarizing Microscope (PM)} manufactured by \textit{Taorad GmbH}, Germany. The microscope has been used in previous 3D-PLI studies to measure the three-dimensional nerve fiber orientations at high resolution \cite{MAxer2011_1,MAxer2011_2,reckfort2015,zeineh2016}. The light source consists of a single white LED (\textit{IntraLED 2020+} operated at 24 W) with integrated K\"ohler illumination and a bandpass filter, generating a wavelength spectrum $\lambda = (550 \pm 5)$\,nm. Further components are a rotatable linear polarizer, a specimen stage, a circular analyzer (quarter-wave retarder combined with linear polarizer), and a CCD camera (monochrome \textit{RETIGA-4000R} camera by \textit{QImaging} with \textit{Kodak KAI-04022-ABA} image sensor) which records an image for each rotation angle $\rho = \{0^{\circ}, 10^{\circ}, \dots, 170^{\circ}\}$ of the polarizer, yielding a series of 18 images. 
The microscope is equipped with a motorized specimen stage (\textit{M\"arzh\"auser}, Germany) which performs a translational scan of the brain section in tiles of $2.7 \times 2.7\,$mm$^2$. To allow for stitching, the tiles were measured with an overlap of 30\% on all sides. The objective lens (\textit{Nikon TL Plan Fluor EPI P 5x}) has a $5\times$ magnification and a numerical aperture of 0.15. The resolution in object space is about 1.33\,\um\,/\,px.

The transmittance and retardation images in \cref{sec:exp-studies,sec:exp-sim} were computed as described in \textsc{Axer} \ea\ \cite{MAxer2011_1,MAxer2011_2} by performing a discrete harmonic Fourier analysis on the measured light intensities $I(\rho)$ per image pixel: $I(\rho) = a_0 + a_2\,\cos(2\rho) + b_2\,\sin(2\rho)$. The transmittance $I_{\text{T}}$ corresponds to the average over all 18 images and was computed from the Fourier coefficient of order zero ($I_{\text{T}} = 2\,a_0$), the retardation $|\sin\delta|$ corresponds to the amplitude of the intensity signal and was computed from the Fourier coefficients of order zero and two $\big(|\sin\delta| = (a_2^2 + b_2^2)^{1/2}/a_0\big)$, where $\delta$ is the phase shift induced by the birefringent brain tissue. The transmittance images were normalized by the transmittance image measured without sample, yielding normalized transmittance images ($I_{\text{T,N}}$).

Images of several consecutive brain sections (see \cref{fig:3D-Transmittance}) were registered onto each other using in-house developed software tools based on the software packages \textit{ITK}, \textit{elastix}, and \textit{ANTs} \cite{elastix,shamonin2013,avants2008,avants2011,itk} which perform linear and non-linear transformations. As undistorted reference volume, aligned blockface images were used: a picture of the brain block surface (\textit{blockface image}) was taken every time before sectioning, a pattern of \textit{ARTag markers} \cite{wagner2007} was used to determine the position of the brain block in two-dimensional space \cite{schober2015}.


\subsection{TPFM measurement}
\label{sec:TPFM}

The TPFM measurements were performed with a custom-made two-photon fluorescence microscope \cite{silvestri2014,costantini2017} at the \textit{European Laboratory for Non-Linear Spectroscopy (LENS)}, University of Florence, Italy. 
The microscope is equipped with a mode-locked titanium-sapphire laser with a wavelength of 800\,nm which is coupled into a scanning system based on a pair of galvanometric mirrors. The laser is focused onto the sample by a water-immersion $25\,\times$ objective lens (\textit{LD LCI Plan-Apochromat 25x/0.8 Imm Corr DIC M27}). The lateral displacement of the sample was realized by a motorized xy-stage (enabling tile-wise scanning of the sample). The axial displacement (along the z-axis) was realized by a closed-loop piezoelectric stage. The fluorescence signals were collected by two photomultiplier tubes, enabling to detect red and green fluorescence.
The setup achieves a resolution of $0.244 \times 0.244 \times 1\,\um^3$. The sample was measured in tiles of $250 \times 250\,\um^2$, with an overlap of 10\% to allow for stitching.


\subsection{Bright-field transmission microscopy}
\label{sec:bright-field}

The bright-field transmission microscopy images were obtained from \textit{ZEISS Axio Imager Vario}. The microscope is equipped with a white microLED which emits unpolarized light with wavelengths between 400\,nm and 750\,nm. The objective lens (\textit{Plan Apochromat 5x}) has a $5\times$ magnification and a numerical aperture of 0.16. The resolution in object space is about 0.91\,\um\,/\,px.


\section{Generation of artificial fiber configurations}
\label{sec:generation-fiber-config}

\subsection{Densely grown fiber bundle}
\label{sec:densely-grown}

\begin{figure*}[!htb]
	\centering
	\includegraphics[width=1.6\columnwidth]{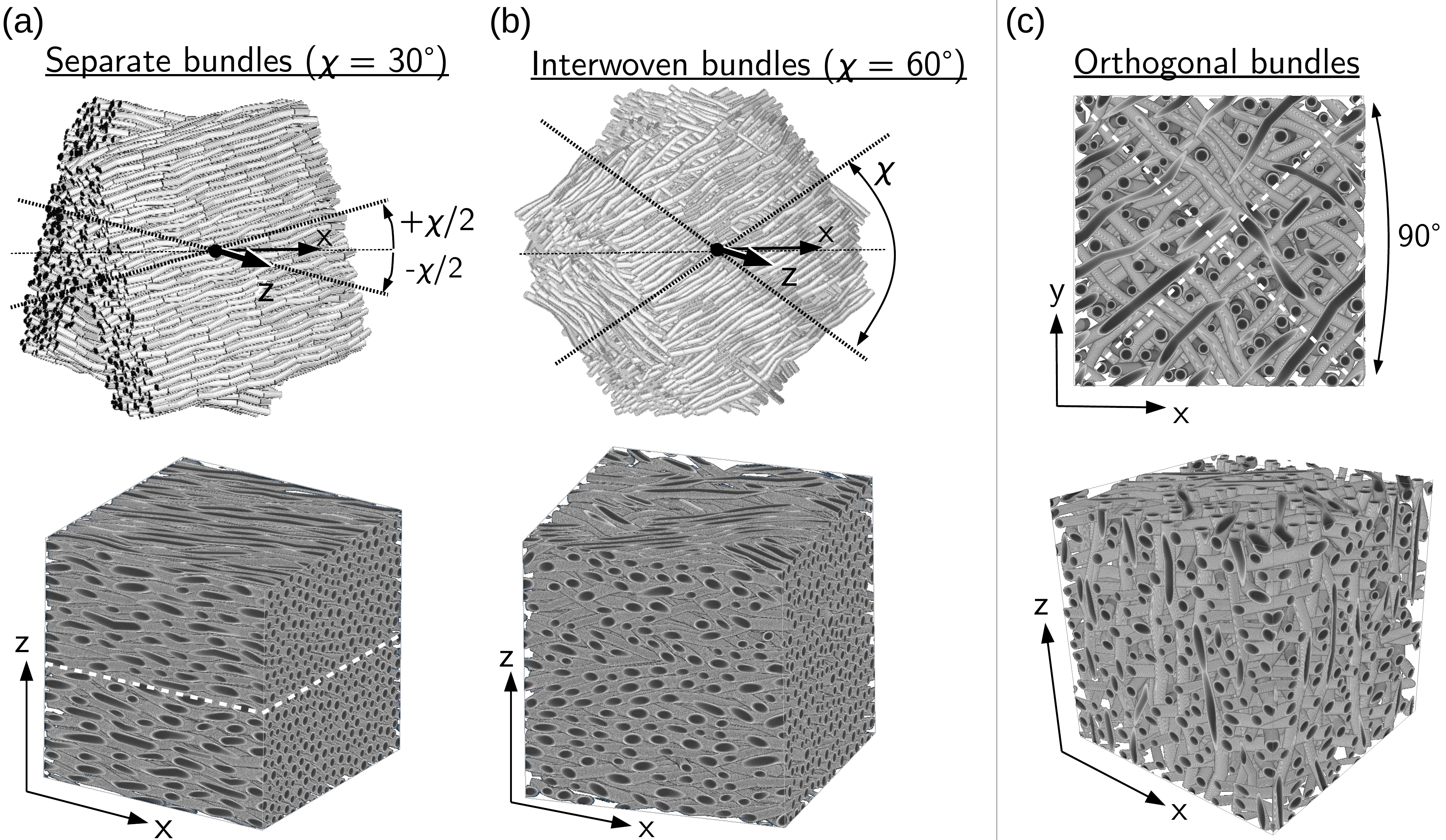}
	\caption{Generation of crossing fibers. (\textbf{a})-(\textbf{b}) Separate and interwoven fiber bundles with crossing angle $\chi$: The upper figures show the generated bundles before cropping. The lower figures show the bundles after being cropped to a volume of $30 \times 30 \times 30\,\um^3$. The white dotted line indicates the border between the upper and the lower bundle of the separate crossing fibers. (\textbf{c}) Three mutually orthogonal, interwoven fiber bundles cropped to a volume of $30 \times 30 \times 30\,\um^3$. The white dotted lines indicate the main directions of the two horizontal fiber bundles in the xy-plane, the third fiber bundle is oriented in the z-direction. All fiber configurations were generated from $700$ fibers with diameters between 1.0\,\um\ and 1.6\,\um.}
	\label{fig:CrossingFibers}
\end{figure*}

The bundle of densely grown fibers (see \cref{fig:Sim_Transmittance_vs_Inclination}(a)) was generated by in-house developed software: $N = 700$ circles with uniformly distributed diameters ($d \in [1.0,1.6]\,\um$) were randomly uniformly placed in the xy-plane (in an area of $45 \times 30\,\um^2$). The circles were initialized with a random speed (max. 0.1\,\um\ displacement per step) and collided with each other (assuming elastic collision with particle mass $r^2$) until a solution was reached without collision in the xy-plane. To obtain well-distributed fibers, the previous step was repeated 250 times before the positions of the circles were stored. To obtain a 3D fiber volume, the circle positions were stored while incrementing the z-position by 1\,\um\ per step.
To generate fiber bundles with different inclination angles, the resulting bundle of densely grown fibers was rotated around the y-axis with respect to the center position and cropped to a volume of $30 \times 30 \times 30 \,\um^3$.
To prevent fibers from touching each other after discretization, all fiber diameters were reduced by 5\%.
In the resulting fiber bundle, about 60\% of the volume is filled with fibers. 


\subsection{Inhomogeneous fiber bundles}
\label{sec:inhom-fiberbundle}

Inhomogeneous fiber bundles, like the bundle with broad fiber orientation distribution (\cref{fig:Sim_Transmittance_vs_Inclination}(c)(ii)) or crossing fibers (\cref{fig:Sim_CrossingFibers}(a),(b)), were generated by in-house developed software \cite{matuschke2019} which allows collision control in 3D. Starting from well-distributed straight fibers with $N = 700$ and $d \in [1.0,1.6]\,\um$ (obtained after 250 steps as described in the previous section), the fibers were divided iteratively into segments of 2--5\,\um\ and assigned a random displacement in the x-, y-, and z-direction. The resulting fiber segments were split or merged until the length of each segment was again between 2--5\,\um, ensuring that the maximum angle between adjacent segments was less than $20^{\circ}$. When a collision between two segments was detected, the segments were exposed to a small repelling force and the previous step was repeated until no more collisions were detected.
To prevent fibers from touching each other, all fiber diameters were reduced by 5\%. The resulting fiber bundle was cropped to a volume of $30 \times 30 \times 30 \,\um^3$.
The fiber bundles were generated from different configurations of straight fibers and different random displacements:
\begin{figure*}[!t]
	\centering
	\includegraphics[width=1.5\columnwidth]{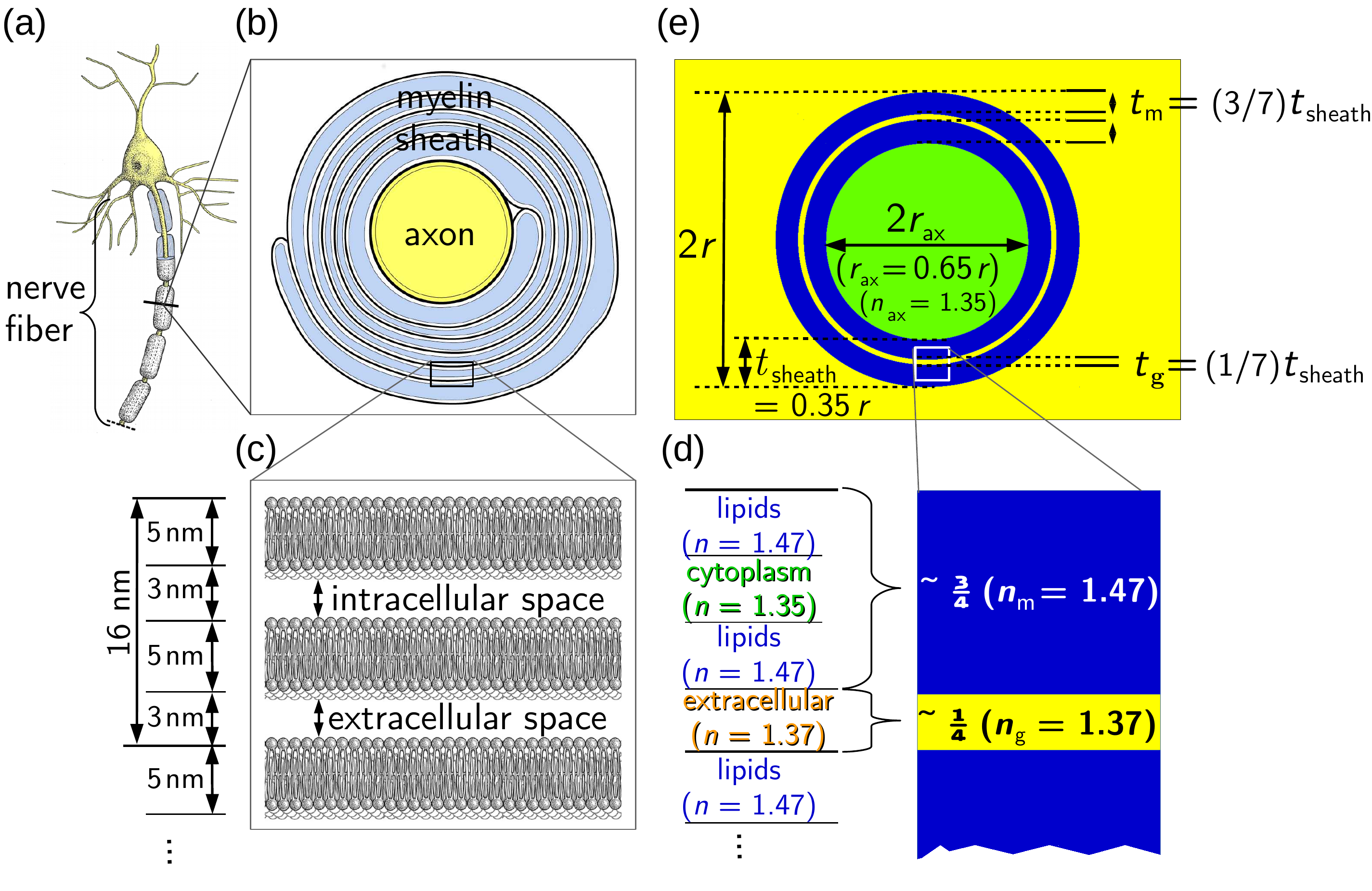}
	\caption{Modeling of nerve fibers. (\textbf{a}) Schematic drawing of a nerve fiber (myelinated axon). (\textbf{b}) Cross-section through the nerve fiber showing the inner axon and the surrounding myelin sheath (formed by a type of glial cell which spirally wraps around the axon). (\textbf{c}) Schematic representation of the myelin structure consisting of several lipid bilayers (5\,nm thick cell membranes) with an intracellular/cytoplasmic and an extracellular space  of about $3\,$nm. (\textbf{d}) Each cell layer (two lipid bilayers with separating cytoplasm) was considered as one ``myelin layer'' with an effective refractive index $\nmyelin = 1.47$ (blue), the extracellular space was considered to be filled with glycerin solution (``glycerin layer'') with a refractive index $n_{\text{g}} = 1.37$ (yellow). The myelin and glycerin layers were assumed to contribute $3/4$ and $1/4$ to the overall myelin sheath thickness $\dmyelin$, respectively. (\textbf{e}) Nerve fibers were modeled with double myelin layers with thickness $t_{\text{m}} = (3/7) \, \dmyelin$ and a single separating glycerin layer with thickness $t_{\text{g}} = (1/7) \, \dmyelin$. The myelin sheath thickness contributes approximately one third to the overall fiber radius ($\dmyelin = 0.35\,r$). The inner axon was modeled with a radius $r_{\text{ax}} = 0.65\,r$ and a refractive index $n_{\text{ax}} = 1.35$.}
	\label{fig:MyelinModel}
\end{figure*}
\begin{itemize}
	\item \textit{Bundle with broad fiber orientation distribution} (\cref{fig:Sim_Transmittance_vs_Inclination}(c)(ii)): The fiber bundle was generated from a bundle of straight horizontal fibers in the x-direction and a maximum random displacement of 10\,\um. In the resulting fiber bundle, about 33\% of the volume is filled with fibers. To generate fiber bundles with different inclination angles, the resulting bundle was rotated around the y-axis with respect to the center position.
	\item \textit{Separate crossing fiber bundles} (\cref{fig:Sim_CrossingFibers}(a)): The bundle of straight horizontal fibers in the x-direction was divided in an upper and a lower bundle of thickness $z/2$, respectively. The upper bundle was rotated around the z-axis about the center position by an angle $+\chi/2$, the lower bundle was rotated by an angle $-\chi/2$, resulting in two separate bundles with crossing angle $\chi$ (cf.\ \cref{fig:CrossingFibers}a). The resulting fibers were used as input for the algorithm with a maximum displacement of 1\,\um. Depending on the crossing angle of the resulting fiber bundle, between 40--50\% of the volume is filled with fibers. 
	\item \textit{Interwoven crossing fiber bundles} (\cref{fig:Sim_CrossingFibers}(a)): Each fiber layer in the z-direction of the straight horizontal fiber bundle (oriented in the x-direction) was rotated alternately by $\pm \chi/2$ (cf.\ \cref{fig:CrossingFibers}(b)). The resulting fibers were used as input for the algorithm with a maximum displacement of 1\,\um. Depending on the crossing angle of the resulting fiber bundles, between 40--50\% of the volume is filled with fibers. 
	\item \textit{Mutually orthogonal, interwoven fiber bundles} (\cref{fig:Sim_CrossingFibers}(b)): The straight horizontal fiber bundle (oriented in the x-direction) was divided in three types of alternating layers: one layer was rotated $+\,45^{\circ}$ around the z-axis, one $-\,45^{\circ}$ around the z-axis, and one $+90^{\circ}$ around the y-axis, yielding two horizontal fiber bundles in the xy-plane and one vertical fiber bundle oriented along the z-axis.
	The resulting fibers were used as input for the algorithm with a maximum displacement of 1\,\um. In the resulting fiber bundle (cf.\ \cref{fig:CrossingFibers}(c)), ca.\ 32\% of the volume is filled with fibers.
\end{itemize}


\section{Model of the nerve fibers}
\label{sec:model-nerve-fibers}

The myelin sheath surrounds most of the axons in the white brain matter and consists of densely packed cell membranes \cite{quarles2006,hildebrand1993}.
\Cref{fig:MyelinModel}(c) shows the layered structure of the myelin sheath: it consists of alternating layers of cell membranes (lipid bilayers of about 5\,nm thickness) and intracellular/cytoplasmic or extracellular space (of about 3\,nm thickness) \cite{martenson,lee2014}. 
As the extracellular membranes are not fused and swell in water \cite{quarles2006,martenson}, it is assumed that the extracellular space is filled with the glycerin solution used for embedding the brain sections (cf.\ \cref{sec:brain-preparation}).

The refractive indices $n$ of the layers were estimated from literature values of lipids/membranes ($n = 1.47$ \cite{vanManen2008}, neglecting any proteins), cytoplasm ($n = 1.35$ \cite{duck1990}), and glycerin solution ($n = 1.37$, measured with digital refractometer).

For the simulation studies in \cref{sec:sim-studies}, a simplified model was used to represent the myelin sheath (see \cref{fig:MyelinModel}(d)): Each cell layer (two lipid bilayers with separating cytoplasm) was considered as one \textit{myelin layer} with an effective refractive index $\nmyelin = 1.47$ (blue), the extracellular space was considered to be filled with glycerin solution (\textit{glycerin layer}) with a refractive index $n_{\text{g}} = 1.37$ (yellow). Assuming that the extracellular space increases when being embedded in glycerin, the myelin and glycerin layers were assumed to contribute $3/4$ and $1/4$ to the overall myelin sheath thickness $\dmyelin$, respectively. The refractive index of the cytoplasmic layer was neglected in this model.

The myelin sheath thickness contributes approximately one third to the overall fiber radius $r$ \cite{morell}. Hence, the myelin sheath thickness was chosen to be $\dmyelin = 0.35\,r$ and the radius of the inner axon $r_{\text{ax}} = 0.65\,r$. The refractive index of the axon (green) was chosen to correspond to the refractive index of cytoplasm ($n_{\text{ax}} = 1.35$). The myelin sheath was modeled as double myelin layers with thickness $t_{\text{m}} = (3/7)\,\dmyelin$ each and a single glycerin layer with thickness $t_{\text{g}} = (1/7)\,\dmyelin$ separating the myelin layers. Interruptions of the myelin sheath (nodes of Ranvier) and the small space between axon and myelin sheath (periaxonal space \cite{hildebrand1993}) were neglected in this model.


\section{FDTD algorithm}
\label{sec:FDTD}

The propagation of the polarized light wave through the brain tissue sample (nerve fiber configuration) was simulated by a massively parallel 3D Maxwell solver based on a conditionally stable finite-difference time-domain (FDTD) algorithm \cite{taflove}. The algorithm computes the electromagnetic field components numerically by discretizing space and time and approximating Maxwell's curl equations by finite differences: The discretization is realized with a cubic Yee grid \cite{yee1966} (each electric field component is surrounded by four magnetic field components and vice versa) and a leapfrog time-stepping scheme. The spatial and temporal derivatives in Maxwell's curl equations are approximated by second-order central differences. For more details, see \textsc{Menzel} \ea\ \cite{menzel2016}.

The simulations were performed with the software \textit{TDME3D}$^{\text{TM}}$ \cite{michielsen2010, wilts2014} -- a massively parallel three-dimensional FDTD Maxwell Solver, Copyright \textit{EMBD} (\textit{European Marketing and Business Development BVBA}).
The software solves Maxwell's equations for arbitrary-shaped objects that are illuminated by arbitrary incident plane waves and that consist of linear, isotropic, lossy materials with known permeability, permittivity, and conductivity.
For the FDTD simulations, a combined algorithmic approach was used: In free space, Yee's algorithm was applied. To compute the interaction of the light with brain tissue, an unconditionally stable \textit{Lie-Trotter-Suzuki product formula approach} was used. This results in a computationally efficient but conditionally stable algorithm. For more information, see \textsc{De~Raedt} \cite{deRaedt}.
The simulations were performed on the supercomputer \textit{JUQUEEN} \cite{juqueen} at Forschungszentrum J\"{u}lich GmbH, Germany.


\section{Simulation parameters}
\label{sec:sim-parameters}

\Cref{tab:Simulation_Parameters} lists the parameters that were used for the simulation studies in \cref{sec:sim-studies}.

\begin{table}[!htb] 
	\caption{Parameters for the simulation studies in \cref{sec:sim-studies}: expenses of one simulation run (computation of one fiber configuration, one wavelength, and one angle of incidence on JUQUEEN), dimensions of the simulation volume, and fiber properties (radius $r$, thickness $t$, refractive index $n$).}
	\label{tab:Simulation_Parameters}
	\centering 
	\begin{tabular}{p{3.7 cm} p{4.5 cm}} 
		\hline\hline\\[0.1pt]
		\multicolumn{2}{l}{\underline{General Simulation Parameters}} \\[3pt] 
		\ \ Yee mesh size: 		& $\varDelta = 25$\,nm \\
		\ \ Courant factor:		& $C = 0.8$ \\
		\ \ \# periods:			& 200 \\
		\ \ MPI grid: 			& $16 \times 16 \times 16$ \\[5pt]
		\ \ core hours: 			&  $\sim 7000$--$8000$ \\
		\ \ wall time: 			&  $\sim$ 1:45--2:00\,h \\ 		
		\ \ min.\ memory required: & $\sim$ 260--$360$\,GB \\[8pt]	
		
		\multicolumn{2}{l}{\underline{Simulation Box}} \\[3pt]
		\ \ volume:				& $x \times y \times z = 30 \times 30 \times 35$ \um$^3$ \\
		\ \ boundaries: 			& UPML ($1\,\um$ thick) \\[8pt]	
		
		\multicolumn{2}{l}{\underline{Surrounding Medium}} \\[3pt]
		\ \ dimensions:			& $x \times y \times z = 30 \times 30 \times 31$ \um$^3$ \\
		\ \ refractive index: 	& $n_{\text{surr}} = 1.37$ \\[8pt]	
		
		\multicolumn{2}{l}{\underline{Fiber Configuration}} \\[3pt]
		\ \ volume:					& $x \times y \times z = 30 \times 30 \times 30$ \um$^3$ \\[3pt]
		\ \ fiber radius: 			& $r \sim 0.5$\,\um \\[3pt]
		\ \ axon: 					& $r_{\text{ax}} = 0.65\,r$,  \,\,\,\,\,\,\,$n_{\text{ax}} = 1.35$ \\[3pt]
		\ \ myelin sheath:  			& $\dmyelin = 0.35\,r = t_{\text{m}} + t_{\text{g}} + t_{\text{m}}$ \\[3pt]
		\ \ double myelin layers: 	& $t_{\text{m}} = \frac{3}{7}\,\dmyelin$, \,\, $\nmyelin = 1.47$ \\[3pt]
		\ \ single glycerin layer: 	& $t_{\text{g}}\, = \frac{1}{7}\,\dmyelin$, \,\,\, $n_{\text{g}} \, = 1.37$ \\[3pt]
		
		\hline\hline
	\end{tabular} 
\end{table}

All fiber configurations were generated in a volume of $30 \times 30 \times 30\,\um^3$.
As described in \cref{sec:model-nerve-fibers}, each fiber was modeled by an inner axon and a surrounding myelin sheath with two layers and different refractive indices (see \cref{fig:MyelinModel}).
The surrounding medium was assumed to be homogeneous with a refractive index $n_{\text{surr}} = n_{\text{g}} = 1.37$, which corresponds to the refractive index of gray brain matter as well as to the refractive index of the surrounding glycerin solution.
To account for the fact that the brain sections are embedded in glycerin solution (see \cref{sec:brain-preparation}), $0.5\,\um$ thick layers of glycerin solution (with refractive index $n_{\text{g}} = 1.37$) were added at the bottom and on top of the sample, yielding a medium with dimensions $30 \times 30 \times 31\,\um^3$.
The dimensions of the simulation box were chosen to be $30 \times 30 \times 35\,\um^3$ to leave some space for light source and detection planes.
The simulation volume was surrounded by uniaxial perfectly matched layer (UPML) absorbing boundaries of 1\,\um\ thickness, thick enough to prevent light from being reflected back into the simulation volume.
The different components of the sample were simulated as dielectrics with real refractive indices (as described in \cref{sec:model-nerve-fibers}).
Absorption was neglected because the absorption coefficients of brain tissue are small \cite{schwarzmaier1997,yaroslawsky2002}.

The simulation studies were performed for a duration of 200 periods and a Courant factor of $0.8$. The Yee mesh size was chosen to be $\varDelta = 25\,$nm. This mesh size is just large enough to account for the double myelin layers of the nerve fiber model (see \cref{sec:model-nerve-fibers}: the glycerin layer for fibers with 1\,\um\ diameter is $25\,$nm thick). 

The light source was modeled as plane monochromatic wave.
The simulation studies were performed for normally incident and coherent light with left-handed circular polarization and a wavelength of 550\,nm (corresponding to the peak wavelength of the employed light source, see \cref{sec:3DPLI}). 
Using an MPI grid of $16 \times 16 \times 16$ on JUQUEEN, each simulation run (\ie, the calculation of one configuration, one wavelength, and one angle of incidence) consumed between 7000--8000 core hours, required a minimum memory between 260--360\,GB, and lasted between 1:45--2:00 hours.


\section{Computation of the transmitted light intensities}
\label{sec:computation-light-intensities}

\begin{figure*}[!t]
	\centering
	\includegraphics[width=1.7\columnwidth]{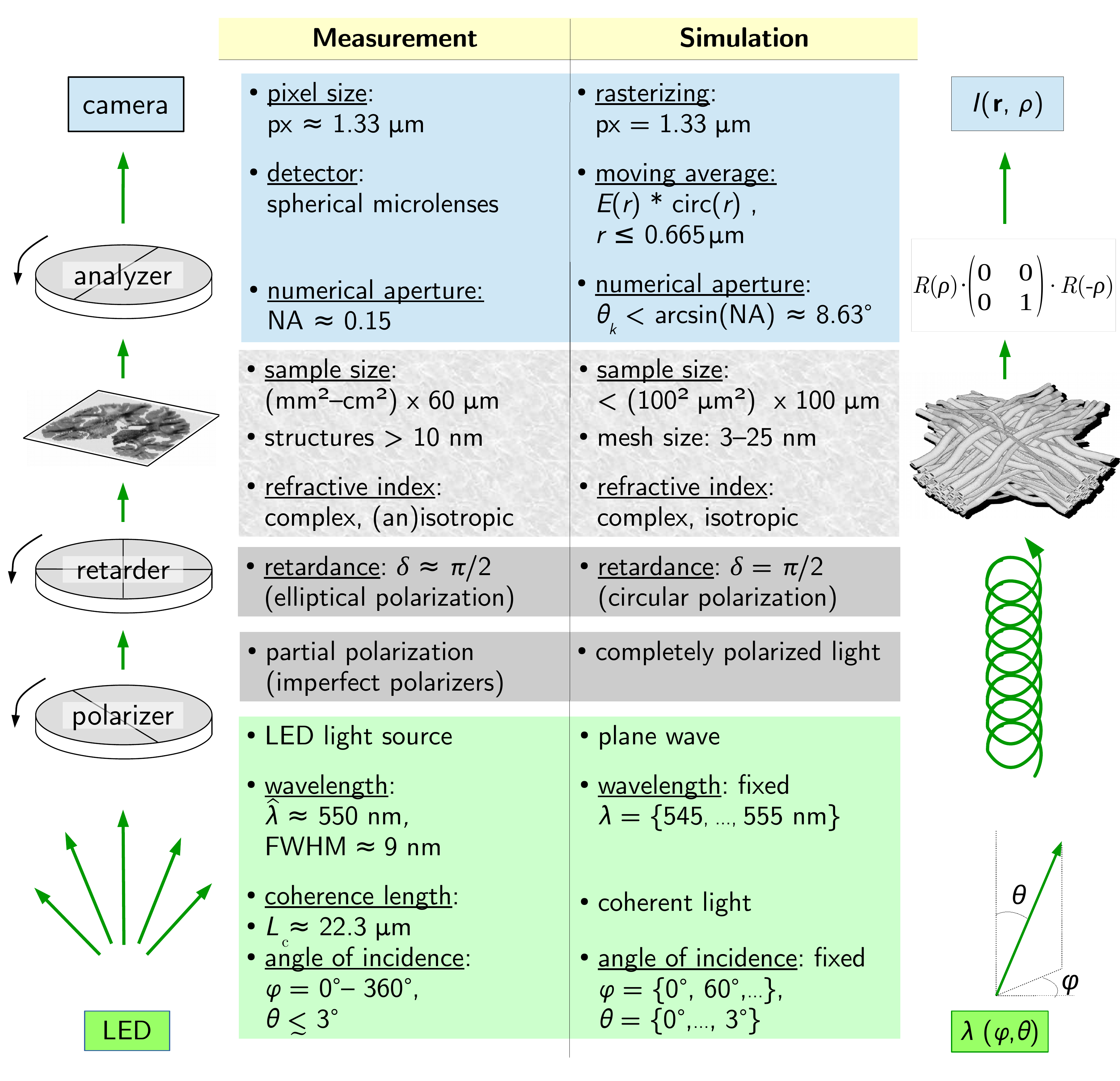}
	\caption{Modeling of the 3D-PLI measurement. The figure and table on the left-hand side show the optical components of the polarimeter (the order of the polarizing filters is different than in the measurement, but the setup is mathematically equivalent): light source (green), polarizer/retarder (dark gray), sample (light gray), objective lens/detector/camera (blue). The table and figure on the right-hand side show how the optical elements were modeled by FDTD simulations: The incoherent and diffusive light source (LED) with peak wavelength $\hat{\lambda}$ and full-width at half-maximum (FWHM) was modeled by performing several simulation runs with plane waves that have different wavelengths ($\lambda$) and angles of incidence ($\varphi$, $\theta$). The modeled light source emits coherent light that is circularly polarized. The tissue sample was represented by an artificial fiber architecture, the rotating analyzer by a rotated Jones matrix (with rotation matrix $R(\rho)$). The numerical aperture (NA) of the imaging system was modeled by considering only wave vector angles $\theta_k < \arcsin(\text{NA})$. The spherical microlenses of the camera detector were modeled by performing a moving average over the area of the microlens with radius $r = 1.33$\,\um\,/\,2.}
	\label{fig:SimulatedPolarimeter}
\end{figure*}

\Cref{fig:SimulatedPolarimeter} shows how the 3D-PLI measurement was modeled by means of FDTD simulations.
For the simulations, a mathematically equivalent polarimetric setup of the employed microscope was considered in which the sample is illuminated by (left-handed) circularly polarized light and analyzed by a rotating linear polarizer (\textit{analyzer}).

The computation of the transmitted light intensities consists of several steps:\\

\noindent \textbf{1.) Maxwell Solver:} 
After passing the polarizing filters in front of the sample (see \cref{fig:SimulatedPolarimeter} on the left), the light wave is left-handed circularly polarized. The propagation of the light wave through the sample was computed by TDME3D as described in \cref{sec:FDTD}. The resulting light wave is represented by a superposition of monochromatic plane waves with different wave vectors $\vec{k}$ and real amplitudes $\vec{E}_{0,k}$:
\begin{align}
\vec{E}_k(\vec{r},t) &= \vec{E}_{0,k} \, \cos(\vec{k}\cdot\vec{r} - \omega t + \phi) \\
&\equiv \vec{A}_k \cos(\vec{k}\cdot\vec{r} - \omega t) - \vec{B}_k \sin(\vec{k}\cdot\vec{r} - \omega t),
\label{eq:PlaneWave_TDME3D_}
\end{align}
where $\vec{r}$ and $t$ are the spatial and temporal coordinates, $\omega$ is the angular frequency, $\phi$ is the phase, and $\vec{A}_k$ and $\vec{B}_k$ are defined as: $\vec{A}_k = \vec{E}_{0,k} \,\cos\phi$ and $\vec{B}_k = \vec{E}_{0,k} \,\sin\phi$.

Note that every index $k$ denotes a different wave vector $\vec{k}$ and is not related to the wave number $k = 2\pi/\lambda$ (the wavelength of the transmitted light waves is the same as for the ingoing light wave).

\vspace{2mm}
\noindent \textbf{2.) Yee shift:}
Before further processing, the electromagnetic field components were shifted in the x,y,z-direction to the middle of the corresponding Yee cell, respectively:
\begin{align}
E_{k,\text{x}}(\vec{r},t): \,\,\,\, y &\mapsto y + \dy/2, \,\,\,\,\,\, z \mapsto z + \dz/2, \label{eq:shift1} \\
E_{k,\text{y}}(\vec{r},t): \,\,\,\, x &\mapsto x + \dx/2, \,\,\,\,\,\, z \mapsto z + \dz/2, \\
E_{k,\text{z}}(\vec{r},t): \,\,\,\, x &\mapsto x + \dx/2, \,\,\,\,\,\, y \mapsto y + \dy/2, \label{eq:shift2}
\end{align}
where $\dx = \dy = \dz$ is the side length of the cubic Yee cell.

For each shift $\Delta j$ in the direction $j = \{x, y, z\}$, the vector components $A_{k,i}$ and $B_{k,i}$ were recomputed as follows:
\begin{align}
\check{A}_{k,i} &= A_{k,i} \,\cos(k_j \, \Delta j) - B_{k,i} \,\sin(k_j \, \Delta j), \\
\check{B}_{k,i} &= A_{k,i} \,\sin(k_j \, \Delta j) + B_{k,i} \,\cos(k_j \, \Delta j).
\end{align}

After performing the shifts specified in \cref{eq:shift1} to (\ref{eq:shift2}), the resulting field vector is given by: 
\begin{align}
\vec{E}'_k(\vec{r},t) &= \vec{A}'_k \cos(\vec{k}\cdot\vec{r} - \omega t) - \vec{B}'_k \sin(\vec{k}\cdot\vec{r} - \omega t).
\end{align}

\vspace{1mm}
\noindent \textbf{3.) Scattering pattern:}	
To study how much light is scattered under a certain angle (wave vector $\vec{k}$), the \textit{scattering pattern} was computed, \ie, the intensity per wave vector normalized by the ingoing light intensity ($I_0$) per image pixel (px):
\begin{align}
I_k \equiv \frac{\vert \vec{E}'_{0,k} \vert^2}{I_0 / (\text{\#\,px})} = \frac{\vert \vec{A}'_k \vert^2 + \vert \vec{B}'_k \vert^2}{I_0 / (\text{\#\,px})}.
\label{eq:diffraction}
\end{align}

\vspace{1mm}
\noindent \textbf{4.) Rotating analyzer:}
To model the 3D-PLI measurement, the electric field vector $\vec{E}'_k(\vec{r},t)$ was processed through the second linear polarizer (analyzer) rotated by angles $\rho$, yielding:
\begin{align}
\vec{\tilde{E}}_k(\vec{r},t,\rho) = \vec{\tilde{A}}_k(\rho) \cos(\vec{k}\cdot\vec{r} - \omega t) - \vec{\tilde{B}}_k(\rho) \sin(\vec{k}\cdot\vec{r} - \omega t).
\label{eq:E_tilde}
\end{align}
The x- and y-components of $\vec{\tilde{E}}_k(\vec{r},t,\rho)$ were computed by multiplying $\vec{E}'_k(\vec{r},t)$ with the Jones matrix of a rotated linear polarizer \cite{jones1941,collett}:
\begin{widetext}
\begin{equation}
\begin{pmatrix} \tilde{E}_{k,\text{x}}(\vec{r},t,\rho)			\\ 
\tilde{E}_{k,\text{y}}(\vec{r},t,\rho)
\end{pmatrix}	= \begin{pmatrix} \cos\rho & -\sin\rho			\\ 
\sin\rho & \cos\rho
\end{pmatrix} \,
\begin{pmatrix} 0\,\, & 0			\\ 
0\,\, & 1
\end{pmatrix} \,
\begin{pmatrix} \cos\rho & \sin\rho			\\ 
-\sin\rho & \cos\rho
\end{pmatrix} \,
\begin{pmatrix} E'_{k,\text{x}}(\vec{r},t)	\\ 
E'_{k,\text{y}}(\vec{r},t)
\end{pmatrix}    			\\	
= 	\begin{pmatrix} \,\,\,\,\,\,\sin\rho \,\, \big(E'_{k,\text{x}}(\vec{r},t) \,\sin\rho - E'_{k,\text{y}}(\vec{r},t) \,\cos\rho\big) \\ 
-\cos\rho \, \big(E'_{k,\text{x}}(\vec{r},t) \,\sin\rho - E'_{k,\text{y}}(\vec{r},t) \,\cos\rho\big)
\end{pmatrix}\,.
\label{eq:E_xy}
\end{equation}
\end{widetext}
The z-component was computed by applying Maxwell's equation in free space and assuming $\vec{\tilde{E}}_k(\vec{r},t,\rho) = \vec{\tilde{E}}_{0,k}(\rho) \, \e^{\I (\vec{k}\cdot\vec{r} - \omega t + \phi)}$ (plane monochromatic wave):
\begin{align}
\text{div} \,\vec{\tilde{E}}_k(\vec{r},t,\rho) &\overset{\hphantom{(\ref{eq:E_xy})}}{=} 0 
\quad \Leftrightarrow \quad \vec{k} \cdot \vec{\tilde{E}}_k(\vec{r},t,\rho) = 0 
\notag\\[5pt]
\Leftrightarrow \,\,
\tilde{E}_{k,\text{z}}(\vec{r},t,\rho) &\overset{\hphantom{(\ref{eq:E_xy})}}{=} - \frac{1}{\kz} \big(\kx \,\tilde{E}_{k,\text{x}}(\vec{r},t,\rho) + \ky \,\tilde{E}_{k,\text{y}}(\vec{r},t,\rho) \big) \notag
\end{align}
\vspace{-0.5cm}
\begin{align}
\overset{(\ref{eq:E_xy})}{=}& - \frac{\kx \sin\rho - \ky \cos\rho}{\kz} \, \notag \\ 
&\times\Big(E'_{k,\text{x}}(\vec{r},t) \, \sin\rho - E'_{k,\text{y}}(\vec{r},t)  \,\cos\rho\Big)\,.
\label{eq:E_z}
\end{align}

\vspace{1mm}
\noindent \textbf{5.) Objective lens:}
The objective lens was assumed to be ideal and both specimen and detector were assumed to lie within the corresponding focal planes of the lens. Thus, the propagation of the electromagnetic wave between sample and detector was assumed to be free and $\vec{\tilde{E}}_k(\vec{r},t,\rho)$ was evaluated at the z-position of the detection plane behind the sample (defined as $z=0$):
\begin{align}
\vec{r} = (r_{\text{x}}, r_{\text{y}}, 0)^{\text{T}}.
\end{align}

To account for the numerical aperture (NA) of the objective lens, only $k$-vectors were processed that fulfill:
\begin{align}
\theta_k = \arccos\left(\frac{\kz}{\sqrt{\kx^2 + \ky^2 + \kz^2}}\right) \,\leq\, \arcsin(\text{NA}).
\end{align}
The employed imaging system has a numerical aperture of about 0.15, so only $k$-vectors with angles $\theta_k \leq 8.6^{\circ}$ were used for processing.

\vspace{2mm}
\noindent \textbf{6.) Detector microlenses:}
The camera sensor contains an array of spherical microlenses which bundle the light onto subjacent photodiodes for each image pixel. 
Assuming perfect microlenses and photodiodes that are completely covered by one microlens, respectively, the microlenses were modeled by applying a moving average over the area of the microlens. Instead of taking the magnification and the physical size of the microlenses into account, the microlenses were modeled with a diameter of $2\,r_0 = 1.33$\,\um\, corresponding to the pixel size of the microscope in object space:
\begin{align} 
\vec{\breve{E}}_k(\vec{r},t,\rho) = \vec{\tilde{E}}_k(\vec{r},t,\rho) * \text{circ}(r) \,\, , \notag \\
\text{circ}(r) = 
\begin{dcases}
\frac{1}{\pi\,r_0^2}\,, \,\, r < r_0 \\
\quad 0\,\,\,, \,\, r \geq r_0.
\end{dcases}
\end{align}

To obtain the full image information (independent of the detector pixel position), no rasterizing was applied.

\vspace{2mm}
\noindent \textbf{7.) Intensity:}
In principle, the intensity detected by the camera sensor depends on the angle of incidence of the incident light: $I\,\cos\theta_k$. As the numerical aperture is sufficiently small (NA $= \sin\theta_k \approx 0.15 \Leftrightarrow \cos\theta_k > 0.9886$), the angle dependence was neglected, which enables to represent the intensity $I(\vec{r},\rho)$ as Fourier series in $\rho$, as described below.

With this assumption, the light intensity recorded by the camera is given by the absolute squared value of the electric field vector. To compute the intensity at a certain point $\vec{r}$ in the image plane, the electric field vectors were summed over $\vec{k}$ and averaged over time:
\begin{align}
I(\vec{r},\rho) &\propto \vert \vec{E}(\vec{r},\rho) \vert^2 \equiv \,\, \frac{1}{T} \int\limits_0^T \Big\vert \sum_{\vec{k}} \vec{\breve{E}}_k(\vec{r},t,\rho) \Big\vert^2 \diff t \notag \\
&\propto \,\,\, \left\vert \iDFT\left\{\vec{\tilde{A}}_k(\rho) + \I \vec{\tilde{B}}_k(\rho) \right\} * \text{circ}(r) \right\vert^2,
\label{eq:I_FT}
\end{align} 
where $\iDFT$ denotes the \textit{inverse discrete Fourier transform}:
\begin{align}
\iDFT \{ f \} = \sum_{\vec{k}} f_{\vec{k}} \, \e^{\I \vec{k}\cdot\vec{r}}.
\end{align}
The \textit{discrete Fourier transform (FT)} is defined analogously.

To save computing time, the convolution in \cref{eq:I_FT} was replaced by a multiplication, making use of the \textit{convolution theorem}: 
\begin{align}
I(\vec{r},\rho) &\propto \Big\vert \iDFT\Big\{ \left( \vec{\tilde{A}}_k(\rho) + \I \vec{\tilde{B}}_k(\rho) \right) \, \DFT\{ \text{circ}(r) \} \Big\} \Big\vert^2 \\
&= \Bigg\vert \iDFT\Bigg\{ \left( \vec{\tilde{A}}_k(\rho) + \I \vec{\tilde{B}}_k(\rho) \right) \, 2\,\frac{J_1(r_0 \, \kxy)}{r_0 \, \kxy}\Bigg\} \Bigg\vert^2,
\end{align}
where the function $J_1(x)$ is the Bessel function of the first kind of order one, with $\kxy \equiv \sqrt{\kx^2 + \ky^2}$ and $r_0 = 0.665\,\um$. 

To simplify notation, the following abbreviations are defined:
\begin{align}
\mathcal{\vec{\tilde{E}}}_k(\rho) &\equiv \left( \vec{\tilde{A}}_k(\rho) + \I \vec{\tilde{B}}_k(\rho) \right) \, 2\,\frac{J_1(r_0 \, \kxy)}{r_0 \, \kxy}, \\
\mathcal{\vec{E}'}_k &\equiv \left( \vec{A}'_k + \I \vec{B}'_k \right) \, 2\,\frac{J_1(r_0 \, \kxy)}{r_0 \, \kxy}, \\
\mathcal{\vec{\tilde{E}}}(\vec{r},\rho) &\equiv \iDFT \big\{\mathcal{\vec{\tilde{E}}}_k(\rho) \big\}, \\
\hspace{0.25cm}\mathcal{\vec{E}'}(\vec{r}) &\equiv \iDFT \big\{\mathcal{\vec{E}}'_k \big\}.
\label{eq:FT_E}
\end{align}

The intensity is then given by:
\begin{align}
I(\vec{r},\rho) \, \propto \,\, \vert \tilde{\mathcal{E}}_{\text{x}}(\vec{r},\rho) \vert^2 + \vert \tilde{\mathcal{E}}_{\text{y}}(\vec{r},\rho) \vert^2 + \vert \tilde{\mathcal{E}}_{\text{z}}(\vec{r},\rho) \vert^2.
\label{eq:I_r_rho}
\end{align}

The x- and y-components of the electric field vector $\vec{\tilde{E}}_k(\vec{r},t,\rho)$ behind the rotating analyzer were computed from $\vec{E}'_k(\vec{r},t) = \vec{A}'_k \cos(\vec{k}\cdot\vec{r} - \omega t) - \vec{B}'_k \sin(\vec{k}\cdot\vec{r} - \omega t)$ according to \cref{eq:E_xy}. As the equation is linear in the x- and y-components of $\vec{E}'_k(\vec{r},t)$, the x- and y-components of \{$\vec{A}'_k$, $\vec{B}'_k$, $\mathcal{\vec{E}'}_k$\} are transformed to \{$\vec{\tilde{A}}_k(\rho)$, $\vec{\tilde{B}}_k(\rho)$, $\mathcal{\vec{\tilde{E}}}_k(\rho)$\} according to the same equation.
As the Fourier transform is independent from $\rho$, \cref{eq:E_xy} also holds for the x- and y-components of $\mathcal{\vec{E}'}(\vec{r})$ and $\mathcal{\vec{\tilde{E}}}(\vec{r},\rho)$, yielding Fourier coefficients of order zero and two:
\begin{widetext}
\begin{align}
&\vert \tilde{\mathcal{E}}_{\text{x}}(\vec{r},\rho) \vert^2 + \vert \tilde{\mathcal{E}}_{\text{y}}(\vec{r},\rho) \vert^2 \notag\\
&\overset{(\ref{eq:E_xy})}{=} \sin^2\rho \, \vert {\mathcal{E}}'_{\text{x}}(\vec{r}) \vert^2 + \cos^2\rho \, \vert {\mathcal{E}}'_{\text{y}}(\vec{r}) \vert^2 
- \, \sin\rho \cos\rho \,\Big( \mathcal{E}'_{\text{x}}(\vec{r}) \, \mathcal{E}'^{\ast}_{\text{y}}(\vec{r}) + \mathcal{E}'^{\ast}_{\text{x}}(\vec{r}) \, \mathcal{E}'_{\text{y}}(\vec{r}) \Big) \notag\\[5pt]
&\overset{\hphantom{(\ref{eq:E_xy})}}{=} \underbrace{\frac{1}{2} \Big( \vert \mathcal{E}'_{\text{x}}(\vec{r}) \vert^2 + \vert {\mathcal{E}}'_{\text{y}}(\vec{r}) \vert^2 \Big)}_{\mathlarger{c_o}} 
+ \underbrace{\frac{1}{2} \Big( \vert \mathcal{E}'_{\text{y}}(\vec{r}) \vert^2 - \vert \mathcal{E}'_{\text{x}}(\vec{r}) \vert^2 \Big) }_{\mathlarger{c_2}} \, \cos(2\rho) 
\underbrace{- \,\frac{1}{2} \Big( \mathcal{E}'_{\text{x}}(\vec{r}) \, \mathcal{E}'^{\ast}_{\text{y}}(\vec{r}) + \mathcal{E}'^{\ast}_{\text{x}}(\vec{r})\, \mathcal{E}'_{\text{y}}(\vec{r}) \Big)}_{\mathlarger{d_2}} \,\sin(2\rho) 
\label{eq:FourierCoefficients_c_d}\\
&\overset{\hphantom{(\ref{eq:E_xy})}}{\equiv} c_0(\vec{r}) + c_2(\vec{r})\, \cos(2\rho) + d_2(\vec{r})\, \sin(2\rho)\,,
\label{eq:coefficients1}
\end{align}
\end{widetext}
where trigonometric identities have been used: $\big(\cos^2 x = \frac{1}{2} + \frac{1}{2} \cos(2x)$, $\sin x \cos x = \frac{1}{2}\sin(2x)\big)$.

Similar analytical calculations yield Fourier coefficients of orders zero, two, and four:
\begin{align}
\vert \tilde{\mathcal{E}}_{\text{z}}(\vec{r},\rho) \vert^2 &= e_0(\vec{r}) + e_2(\vec{r})\, \cos(2\rho) + f_2(\vec{r})\, \sin(2\rho) \notag\\
& \quad + e_4(\vec{r})\, \cos(4\rho) + f_4(\vec{r})\, \sin(4\rho)\,,
\label{eq:E_z_Fourier}
\end{align}
where $e_m(\vec{r})$ and $f_m(\vec{r})$ are functions of the inverse discrete Fourier transforms:
\begin{align}
X_{\text{x}}(\vec{r}) &\equiv \iDFT \left\{ \frac{\kx}{\kz} \mathcal{E}'_{k,\text{x}} \right\}\, , 
&X_{\text{y}}(\vec{r}) \equiv \iDFT \left\{ \frac{\ky}{\kz} \mathcal{E}'_{k,\text{x}} \right\}\, ,
\label{eq:Xx_Xy}\\
Y_{\text{x}}(\vec{r}) &\equiv \iDFT \left\{ \frac{\kx}{\kz} \mathcal{E}'_{k,\text{y}} \right\}\, , 
&Y_{\text{y}}(\vec{r}) \equiv \iDFT \left\{ \frac{\ky}{\kz} \mathcal{E}'_{k,\text{y}} \right\}\,.
\label{eq:Yx_Yy}
\end{align}

Thus, the transmitted light intensity $I(\vec{r},\rho)$ can be written in terms of a Fourier series:
\setlength{\abovedisplayskip}{10pt}
\begin{align}
I(\vec{r},\rho) \, &\propto \,\, \vert \tilde{\mathcal{E}}_{\text{x}}(\vec{r},\rho) \vert^2 + \vert \tilde{\mathcal{E}}_{\text{y}}(\vec{r},\rho) \vert^2 + \vert \tilde{\mathcal{E}}_{\text{z}}(\vec{r},\rho) \vert^2 \notag \\[5pt]
&= \,\, a_0(\vec{r}) + a_2(\vec{r}) \, \cos(2\rho) + b_2(\vec{r}) \, \sin(2\rho) \notag \\
&\quad\quad + a_4(\vec{r}) \, \cos(4\rho) + b_4(\vec{r}) \, \sin(4\rho), 
\label{eq:I_PLI}\\[5pt]
a_0(\vec{r})  &\equiv c_0(\vec{r})  + e_0(\vec{r}) , \,\,\,\,\, a_2(\vec{r})  \equiv c_2(\vec{r})  + e_2(\vec{r}) , \notag \\
b_2(\vec{r})  &\equiv d_2(\vec{r})  + f_2(\vec{r}) , \,\,\,\,\, a_4(\vec{r}) \equiv e_4(\vec{r}) , \,\,\,\,\, b_4(\vec{r})  \equiv f_4(\vec{r}) \,,
\label{eq:Maxwell_FourierCoefficients}
\end{align}
where the Fourier coefficients $a_m(\vec{r})$ and $b_m(\vec{r})$ are computed from the six inverse discrete Fourier transforms defined above: $\mathcal{E}'_{\text{x}}(\vec{r})$, $\mathcal{E}'_{\text{y}}(\vec{r})$, $X_{\text{x}}(\vec{r})$, $X_{\text{y}}(\vec{r})$, $Y_{\text{x}}(\vec{r})$, $Y_{\text{y}}(\vec{r})$.

For non-normally incident light ($\kx \neq 0$ or $\ky \neq 0$), the transmitted light intensity contains Fourier coefficients of order four (cf.\ \cref{eq:E_z_Fourier}).


Using \cref{eq:I_PLI}, the light intensity was computed for arbitrary rotation angles $\rho$ and normalized by the ingoing light intensity per image pixel:
\begin{align}
I_{\text{N}}(\vec{r},\rho) = \frac{I(\vec{r},\rho)}{I_0 / (\text{\#\,px})}.
\label{eq:INorm}
\end{align}
In the experiment, the measured light intensities are normalized by the light intensities measured without specimen to compensate for filter inhomogeneities. This image calibration could be modeled by performing an additional simulation run without sample. To save computing time, the simulated light intensities were simply normalized by $I_0$ (without considering the imaging system) and only relative values were used for the comparison between measured and simulated light intensities.

The Fourier coefficient of order zero $a_{0,\text{N}}(\vec{r})$ obtained from the simulated normalized transmitted light intensity $I_{\text{N}}(\vec{r},\rho)$ was used to compute the simulated transmittance images $I_{\text{T,N}}(\vec{r})$:
\begin{align}
I_{\text{T,N}}(\vec{r}) \equiv a_{0,\text{N}}(\vec{r}).
\end{align}


Figure S\ref{fig:FlowChart} in the Supplemental Material summarizes the most important steps of computing the transmitted light intensities for 3D-PLI simulations.
The computation was carried out in \textsc{Python} (version 2.7.6) using the \textit{NumPy} package (version 1.12.1) \cite{oliphant2007,walt2011}. To obtain the intensity at a certain pixel position (x,y), the inverse discrete Fourier transform was computed in two dimensions by means of the \textit{Fast Fourier Transform} \cite{cooley1965}.
To enable an efficient use of the FFT, the number of grid points in x and y ($N_{\text{x}}$ and $N_{\text{y}}$) were set to be a multiple of two:
\begin{align}
N'_{\text{x}} &= 2^{m_{\text{x}}} > N_{\text{x}}, \\
N'_{\text{y}} &= 2^{m_{\text{y}}} > N_{\text{y}}.
\end{align}


\section{Error estimation of simulation results}
\label{sec:error-estimation}

When modeling the optical components of the imaging system, the limitations of the simulation software need to be taken into account: the simulated light wave is completely polarized and coherent, the materials are characterized by isotropic refractive indices, and size and resolution of the simulated geometries are limited due to finite computing time.

Using completely polarized light for the simulations implies that the optical elements are assumed to be ideal (unpolarized light source, ideal polarizing filters, no polarization-sensitivity of the camera). For the employed polarizing microscope, these assumptions are reasonable because the optical components are of high quality. Moreover, the transmittance can be considered to be mostly independent from the polarization properties of the imaging system.

\begin{figure*}[!t]
\centering
\includegraphics[width=1.4\columnwidth]{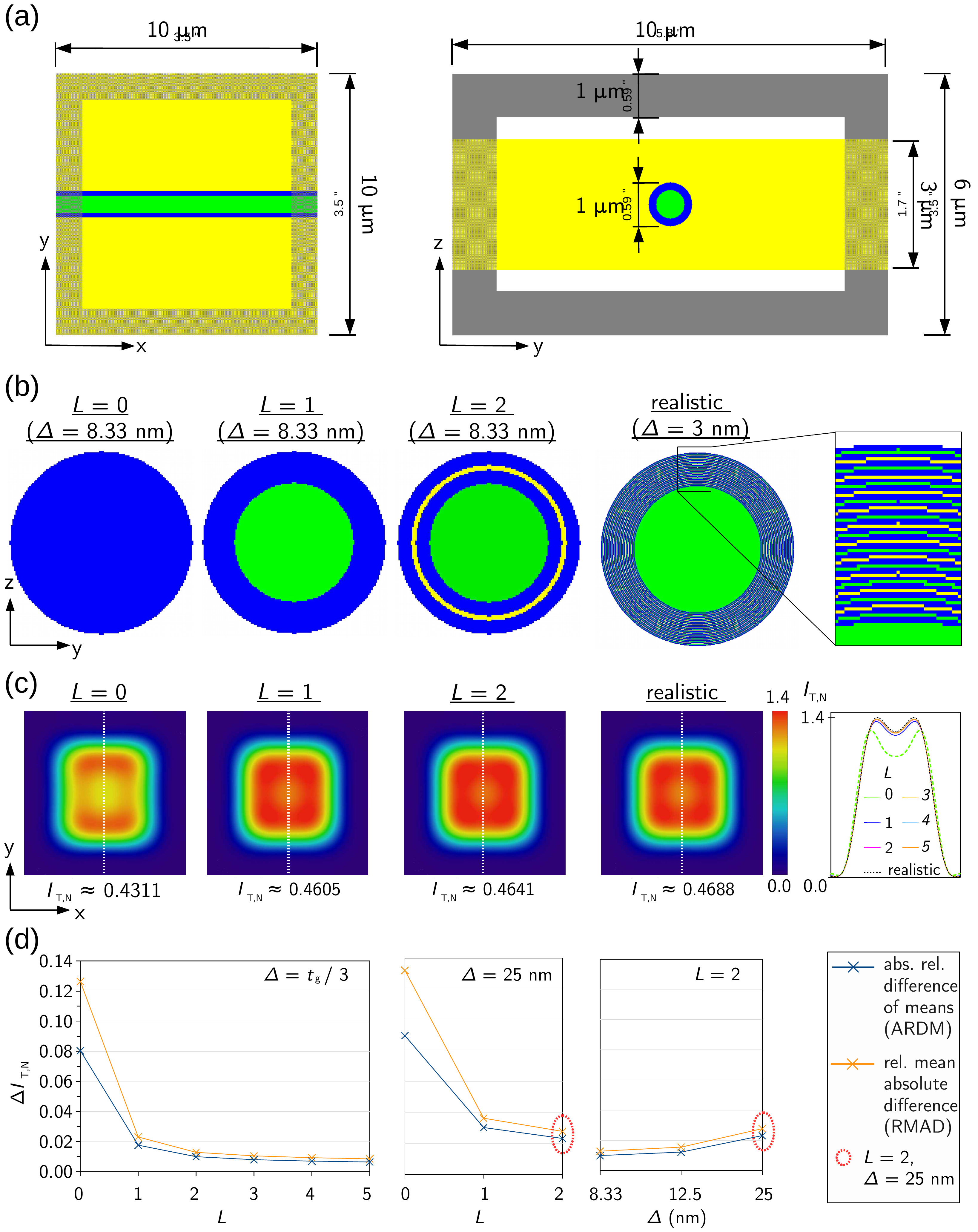}
\caption{Error estimation for different numbers of myelin layers. (\textbf{a}) Dimensions of the simulation volume (xy/yz-plane) used to simulate a straight single fiber with different numbers of myelin layers. (\textbf{b}) Cross-section through fibers with different numbers $L$ of myelin layers and different Yee mesh sizes $\varDelta$. All fibers were modeled with a diameter of $1\,\um$, consisting of an inner axon (green) with a diameter of $0.65\,\um$ and a surrounding myelin sheath with a thickness of $0.175\,\um$. The myelin sheath is composed of alternating layers of myelin (blue) and glycerin (yellow), the myelin layers are three times thicker than the glycerin layers. The realistic model of the myelin sheath contains 22 layers of 5\,nm thick cell membranes (blue), interrupted by 3\,nm thick alternating layers of cytoplasm (green) and surrounding glycerin solution (yellow), yielding a myelin sheath composed of 43 thin layers. The refractive indices are 1.35 for the axon/cytoplasm (green), 1.37 for the glycerin solution (yellow), and 1.47 for the myelin layers (blue). A motivation of the myelin sheath model and the corresponding refractive indices is shown in \cref{fig:MyelinModel}. (\textbf{c}) Normalized transmittance images, corresponding mean values $\overline{I_{\text{T,N}}}$, and transmittance profiles on the right (middle image pixels evaluated along the y-axis, see white dashed lines) obtained from 3D-PLI simulations with different numbers $L$ of myelin layers and Yee mesh sizes defined in (b). The simulations were performed for normally incident light with 550\,nm wavelength and simulation parameters specified in \cref{tab:Simulation_Parameters}. The profiles with non-italic labels belong to the displayed transmittance images. (\textbf{d}) Relative differences between the transmittance images with different numbers $L$ of myelin layers and different mesh sizes $\varDelta$ (relative to the glycerin layer thickness $t_{\text{g}}$) and the transmittance image with realistic myelin sheath. The values for ARDM (blue) and RMAD (orange) were computed using \cref{eq:ARDM} and \cref{eq:RMAD}, the values surrounded in red belong to fibers with double myelin layers ($L=2$) and 25\,nm mesh size, which were used for the simulation studies in \cref{sec:sim-studies}.}
\label{fig:SimParameters_MyelinLayers}
\end{figure*}

The simulation studies in \cref{sec:sim-studies} were performed for a reduced sample size ($30 \times 30 \times 30\,\um^3$) and 200 periods. Simulations with larger sample sizes in x/y and more periods yielded similar results \cite{MMenzel}.

To further reduce computing time, the simulation studies were performed for a simplified nerve fiber model (axon surrounded by double myelin layers, cf.\ \cref{fig:MyelinModel}(e)), a Yee mesh size of 25\,nm, and normally incident light with 550\,nm wavelength.
To estimate the accuracy of the simulation results, the transmittance images were simulated for different numbers of myelin layers $L$, different Yee mesh sizes $\varDelta$, different wavelengths $\lambda$, and different angles of incidence $\theta$. To study the influence of one simulation parameter at once, only one simulation parameter was varied while all other simulation parameters were chosen as in \cref{tab:Simulation_Parameters} (with normally incident light and 550\,nm wavelength).

To estimate the accuracy of the resulting transmittance images, the \textit{absolute relative difference between the mean values (ARDM)} and the \textit{relative mean absolute difference (RMAD)} between the images were computed:

\begin{align}
\text{ARDM} &\equiv \left\vert \frac{\langle \text{image} \rangle - \langle \text{ref.image} \rangle}{\langle \text{ref.image} \rangle} \right\vert, 
\label{eq:ARDM} \\
\text{RMAD} &\equiv \frac{\langle \vert \text{image} - \text{ref.image} \vert \rangle}{\vert\langle\text{ref.image}\rangle\vert}.
\label{eq:RMAD}
\end{align}
\begin{figure*}[!htb]
	\centering
	\includegraphics[width=1.5\columnwidth]{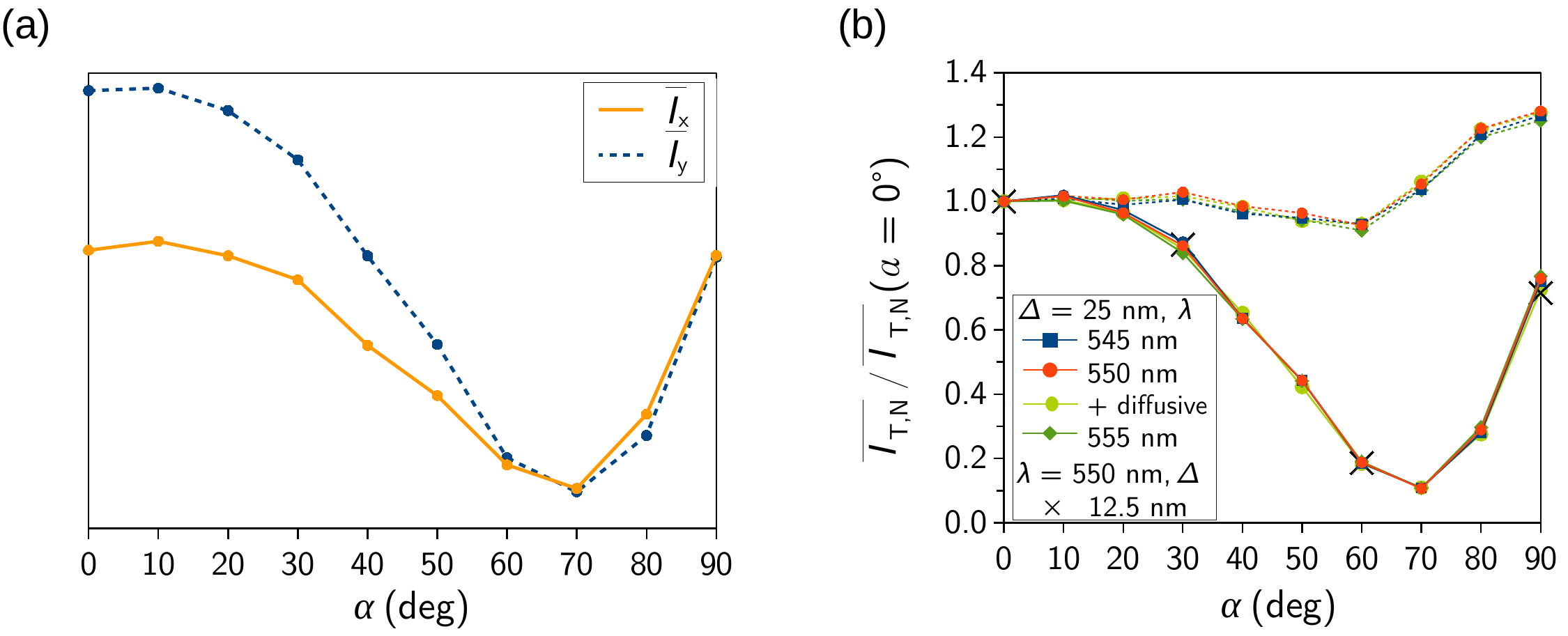}
	\caption{Simulated transmitted light intensity for different simulation parameters. Bundle of densely grown fibers (cf.\ \cref{fig:Sim_Transmittance_vs_Inclination}(a)) simulated for different inclination angles $\alpha$: (\textbf{a}) Mean transmitted light intensity for light polarized along the x-axis ($\overline{I_{\text{x}}}$) or along the y-axis ($\overline{I_{\text{y}}}$) plotted against $\alpha$. The simulations were performed for a numerical aperture NA = 0.15 using a normally incident plane wave with 550\,nm wavelength and simulation parameters specified in \cref{tab:Simulation_Parameters}. (\textbf{b}) Transmittance curves (mean transmittance value $\overline{I_{\text{T,N}}}$ plotted against $\alpha$) obtained from 3D-PLI simulations for normally incident light with different wavelengths $\lambda = \{545, 550, 555\}\,$nm and a Yee mesh size of $\varDelta = 25$\,nm. The simulations for $\lambda = 550$\,nm were also performed for diffusive light (with angles of incidence $\{\theta = 0^{\circ}\}$, $\{\theta = 3^{\circ}$; $\varphi = 4 \times 90^{\circ}\}$). The curves were normalized by the mean transmittance value of the horizontal bundle, respectively. The solid curves were computed for the numerical aperture of the imaging system (NA = 0.15), the dashed curves were computed for NA~=~1. The black crosses belong to simulations with NA = 0.15, $\lambda = 550$\,nm, and $\varDelta = 12.5$\,nm.}
	\label{fig:SimParameters_TransmittanceCurves}
\end{figure*}
In this notation, the ``image'' refers to the transmittance image for which the absolute relative difference is computed (obtained, \eg, from simulations with different Yee mesh sizes). The ``reference image'' is the transmittance image used for comparison (obtained, \eg, from the simulation with minimum mesh size). The symbol $\langle \rangle$ represents the average over all image pixels. As the simulation studies mostly investigate the mean transmittance values, the ARDM is a direct measure for the accuracy of the simulation results, while the RMAD is a measure for the reliability of the ARDM as an error estimate.


\subsection{Different numbers of myelin layers}
\label{sec:diff-myelinlayers}

To estimate the accuracy of the simplified nerve fiber model, a straight single fiber with reduced simulation volume (see \cref{fig:SimParameters_MyelinLayers}(a)) was simulated for different numbers $L$ of myelin layers with thickness $t_{\text{m}}$ (and $L-1$ separating glycerin layers with thickness $t_{\text{g}} = t_{\text{m}}/3$) as well as for a realistic model of the myelin sheath consisting of 43 thin layers (see \cref{fig:SimParameters_MyelinLayers}(b)). 
The Yee mesh size was chosen to be small enough to resolve all geometric features: For most samples, the mesh size was chosen to be one third of the glycerin layer thickness ($\varDelta = t_{\text{g}}/3$). Fibers with two myelin layers ($L=2$) were also simulated for larger mesh sizes ($\varDelta = t_{\text{g}}/2 = 12.5$\,nm and $\varDelta = t_{\text{g}} = 25\,$nm). The realistic myelin sheath was simulated for $\varDelta = t_{\text{g}} = 3$\,nm, consuming 288\,358 core hours on JUQUEEN (using an MPI grid of $64 \times 64 \times 16$).

\Cref{fig:SimParameters_MyelinLayers}(c) shows the corresponding transmittance images, mean values, and line profiles obtained from 3D-PLI simulations with normally incident light and $\lambda = 550$\,nm for the straight single fibers shown in \cref{fig:SimParameters_MyelinLayers}(b). The mean values and line profiles for $L \geq 1$ look similar.
For better comparison, \cref{fig:SimParameters_MyelinLayers}(d) shows the absolute relative differences (ARDM and RMAD) between the transmittance images with $L = \{0,1,2,3,4,5\}$ and the transmittance image with realistic myelin sheath. The relative differences decrease with increasing number of myelin layers $L$ and with decreasing mesh size $\varDelta$.  
A fiber with two or more myelin layers and a mesh size $\varDelta = t_{\text{g}}/3$ yields similar transmittance values as the fiber with realistic myelin sheath.
With increasing mesh size, the relative differences increase. For a fiber with double myelin layers and a mesh size $\varDelta = 12.5$\,nm ($25$\,nm), the differences are: ARDM $\approx$ 1.2\% (2.3\%) and RMAD $\approx$ 1.6\% (2.8\%). For a mesh size of 25\,nm, the differences are still smaller than for a fiber without or with a single myelin layer. 
Thus, a fiber with double myelin layers and a mesh size of 25\,nm is a good compromise between accuracy and computing time and was used for all simulation studies in \cref{sec:sim-studies}. In interesting cases, the simulations were repeated for a reduced mesh size ($\varDelta = 12.5$\,nm), see black crosses in \cref{fig:SimParameters_TransmittanceCurves}(b).


\subsection{Different wavelengths, angles of incidence, \\and Yee mesh sizes}
\label{sec:diff-wavelengths}

The light source of the employed Polarizing Microscope emits light with slightly different wavelengths ($\lambda = (550 \pm 5)$\,nm) and different angles of incidence (the sample is illuminated under angles $\theta < 3^{\circ}$) \cite{MMenzel}. 
To model this incoherent and diffusive light source, several simulation runs with different wavelengths $\lambda$ and angles of incidence ($\varphi$, $\theta$) were performed, and the resulting intensities were added incoherently.
A comparison of simulated and experimental data for a well-defined sample (\textit{USAF-1951} resolution target) revealed that the light source can sufficiently be modeled by three different wavelengths ($\lambda = \{545, 550, 555\}$\,nm) weighted according to the wavelength spectrum, and five angles of incidence ($\theta=0^{\circ}$; $\theta = 3^{\circ}$, $\varphi = \{0^{\circ}, 90^{\circ}, 180^{\circ}, 270^{\circ}\}$) \cite{MMenzel}.

The simulation studies in \cref{sec:sim-studies} were only performed for normally incident light and a single wavelength ($\lambda = 550$\,nm).
To estimate the accuracy of the simulation results, especially for the transmittance curves in \cref{fig:Sim_Transmittance_vs_Inclination}(c), the 3D-PLI simulations for the bundle of densely grown fibers with inclination angles $\alpha = \{0^{\circ}, 10^{\circ}, \dots, 90^{\circ}\}$ were performed for the three different wavelengths and five angles of incidence defined above. The resulting transmittance curves for NA~=~0.15 (solid curves) and NA = 1 (dashed curves) are shown in \cref{fig:SimParameters_TransmittanceCurves}(b). The simulations were performed for a Yee mesh size of 25\,nm. For some inclination angles, the simulations for NA = 0.15 were repeated for a smaller mesh size ($\varDelta = 12.5$\,nm), see black crosses.

The transmittance curves for different wavelengths and for diffusive light (obtained from simulation runs with different angles of incidence) look all very similar. The maximum difference between the normalized transmittance values is less than 0.03. In addition, the simulations with smaller mesh size (black crosses) yield similar results as the simulations with larger mesh size (curves), the maximum difference between the normalized transmittance values is only about 0.005. 
Thus, the transmittance curves for the bundle of densely grown fibers are not sensitive to small changes in wavelength, angle of incidence, or mesh size, which demonstrates the validity of the simulation results.




\bibliography{BIBLIOGRAPHY}


\newpage


\begin{figure*}[!t]
	\centering
	\includegraphics[width=1.3\columnwidth]{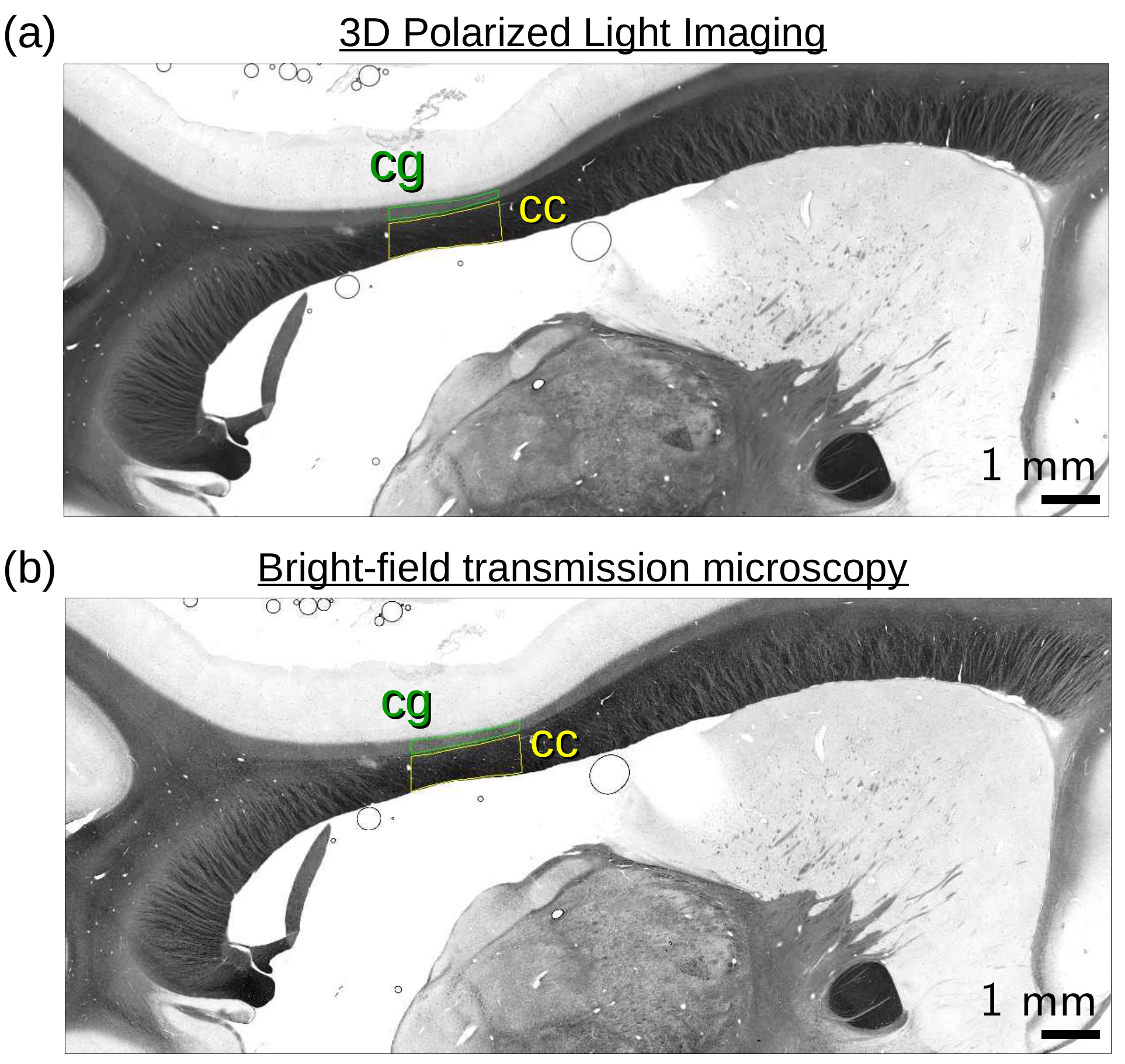}
	\caption{\textbf{Supplementary Figure $\vert$} 3D-PLI measurement vs. bright-field transmission microscopy of a sagittal vervet brain section (neighboring section to the one shown in \cref{fig:Rat_Vervet_Atlas}(b), enlarged region): (\textbf{a}) Normalized transmittance image obtained from a 3D-PLI measurement (see \cref{sec:3DPLI}). (\textbf{b}) Bright-field transmission microscopy image obtained from \textit{ZEISS Axio Imager Vario} with unpolarized light (see \cref{sec:bright-field}). The corpus callosum (cc) contains mainly out-of-plane nerve fibers, while the cingulum (cg) contains mainly in-plane nerve fibers (cf.\ \cref{fig:Rat_Vervet_Atlas}). The mean transmitted light intensity of the selected region in the corpus callosum is about 53\% (a) and 32\% (b) less than the mean transmitted light intensity of the selected region in the cingulum, respectively. This demonstrates that the same effects (less transmitted light intensity in regions with out-of-plane nerve fibers than in regions with in-plane nerve fibers) can also be observed in conventional transmission microscopy which uses unpolarized light.}
	\label{fig:Vervet_PLI-vs-Zeiss}
\end{figure*}


\begin{figure*}[!t]
	\centering
	\includegraphics[width=1.3\columnwidth]{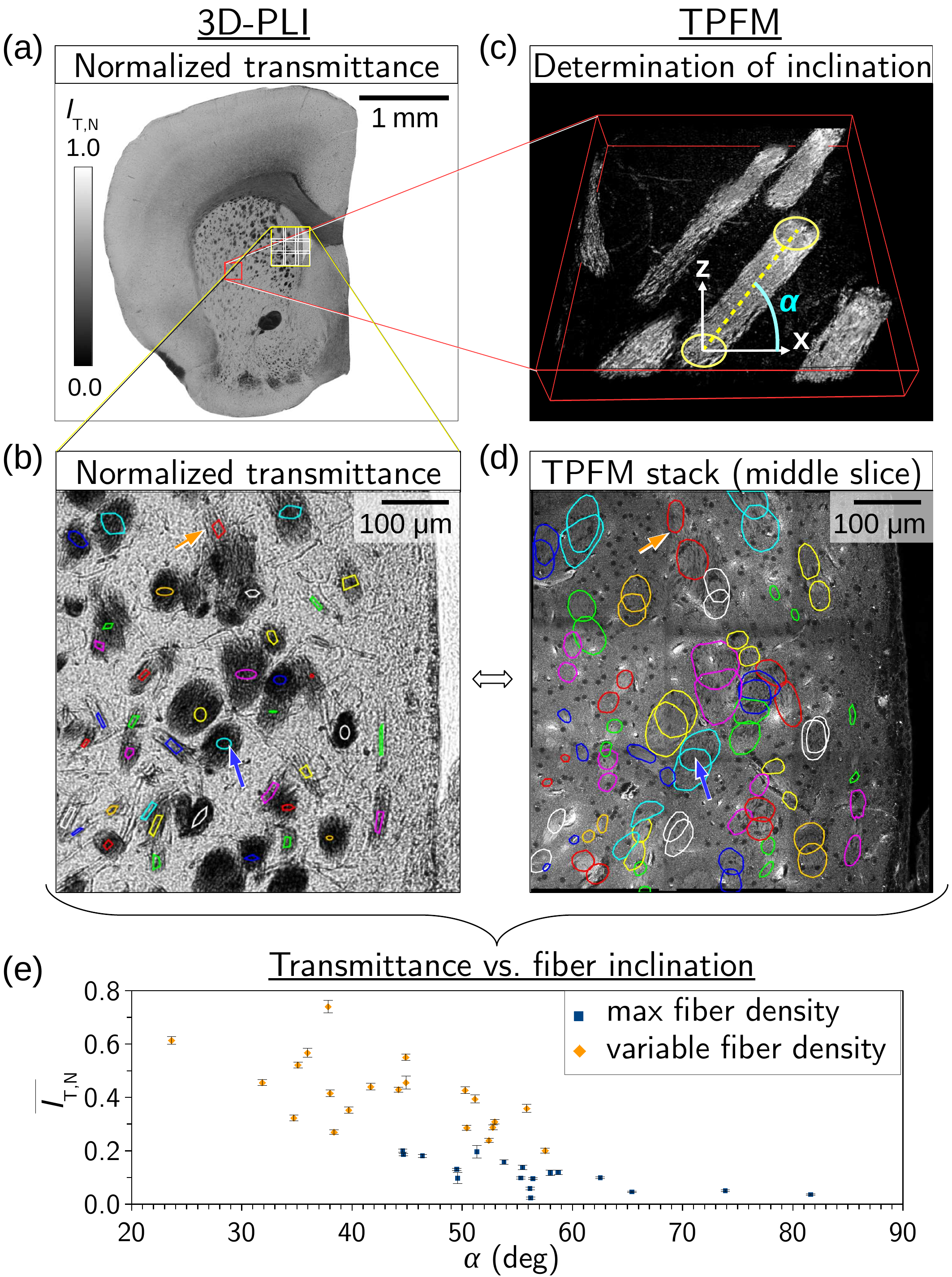}
	\caption{\textbf{Supplementary Figure} $\vert$ Transmittance vs.\ inclination of nerve fiber bundles. A coronal, 60\,\um\ thin section from the left hemisphere of a mouse brain was measured with 3D-PLI (voxel size: $1.33 \times 1.33 \times 60\,\um^3$) and with TPFM (voxel size: $0.24 \times 0.24 \times 1\,\um^3$) for a region of 3 $\times$ 3 tiles in the caudate putamen:
		(\textbf{a}) normalized transmittance image of the coronal mouse brain section, (\textbf{b}) normalized transmittance image of the selected region in the caudate putamen, (\textbf{c}) single tile of a contrast-enhanced TPFM image stack demonstrating the determination of the fiber inclination angle $\alpha$, (\textbf{d}) middle slice of the TPFM image stack for the selected region in the caudate putamen, (\textbf{e}) mean normalized transmittance values $\overline{I_{\text{T,N}}}$ plotted against the determined fiber inclination angles $\alpha$. The inclination angles of 40 fiber bundles were determined from the center point positions of the cross-sections in the first and the last slice of the TPFM image stack (see colored shapes in Fig.\ (d)), the transmittance values were evaluated in the middle of the corresponding fiber bundles (see colored shapes in Fig.\ (b)). The blue (orange) data points in the scatter plot belong to regions that are (are not) completely filled with nerve fibers, cf.\ blue (orange) arrows in Figs.\ (b),(d). The error bars indicate the standard error of the mean for the evaluated transmittance values for each region. Although the values are broadly distributed, the scatter plot shows a clear tendency towards a decrease in transmittance with increasing fiber inclination, also in regions with maximum fiber density.}
	\label{fig:PLI_vs_TPFM}
\end{figure*}


\begin{figure*}[!t]
	\centering
	\includegraphics[width=1.7\columnwidth]{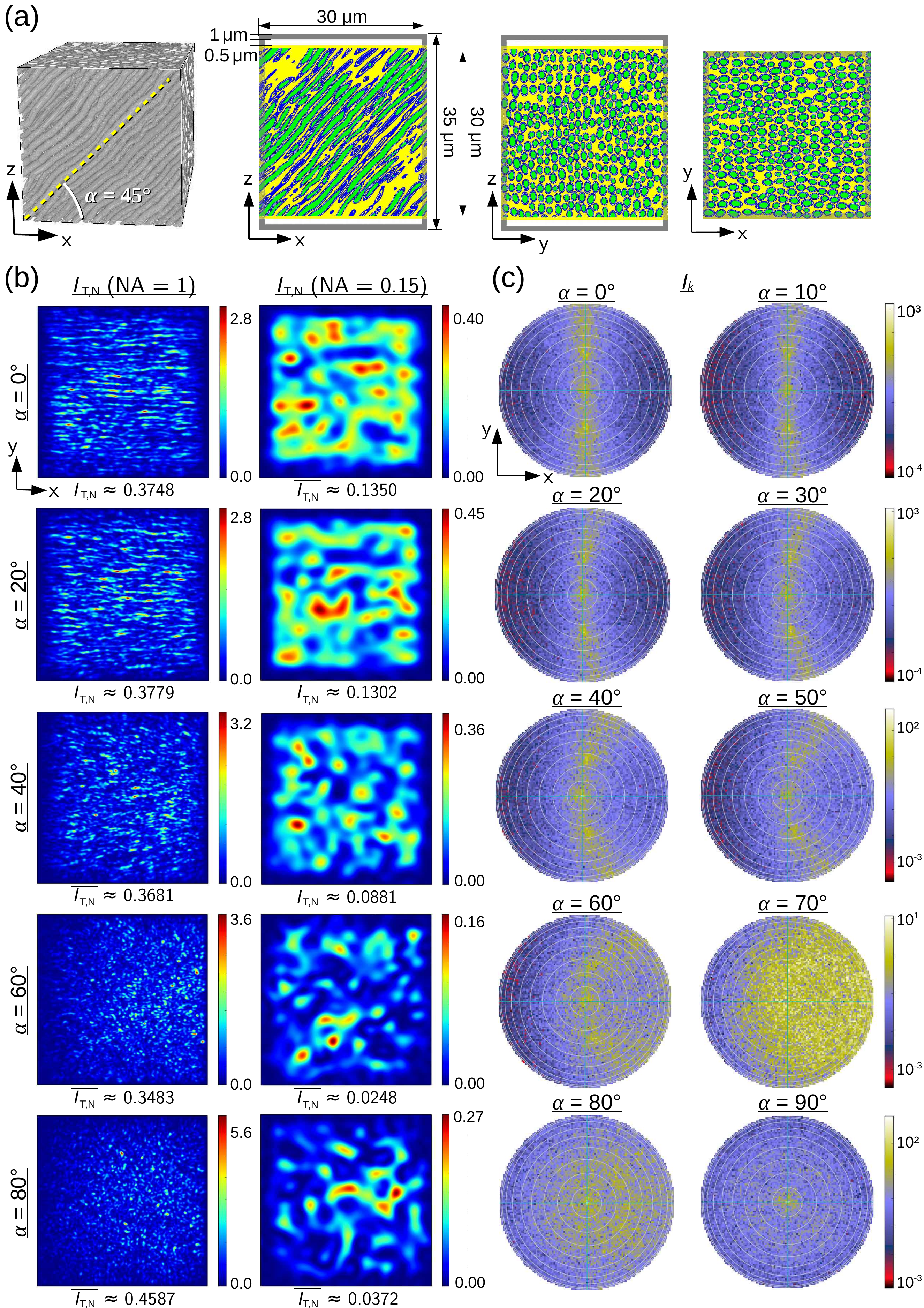}
	\caption{\textbf{Supplementary Figure} $\vert$ Simulated transmittance and light-scattering patterns for different inclinations of the densely grown fiber bundle. (\textbf{a}) 3D-view and cross-sections through mid-planes of the simulation volume for the  densely grown fiber bundle with inclination $\alpha = 45^{\circ}$. (\textbf{b})-(\textbf{c}) Transmittance images $I_{\text{T,N}}$ (left) and light-scattering patterns $I_k$ (right) computed from a simulated 3D-PLI measurement for different inclination angles $\alpha$ of the fiber bundle (using a normally incident plane wave with 550\,nm wavelength and simulation parameters specified in \cref{tab:Simulation_Parameters}). The transmittance images were computed for the numerical aperture of the imaging system (NA = 0.15) and for NA = 1. Note that the value ranges differ between images. With increasing fiber inclination, the scattering increases which leads to a decrease of the mean transmittance values $\overline{I_{\text{T,N}}}$ for NA = 0.15.}
	\label{fig:Sim_TransmittanceImages}
\end{figure*}


\begin{figure*}[!t]
	\centering
	\includegraphics[width=2\columnwidth]{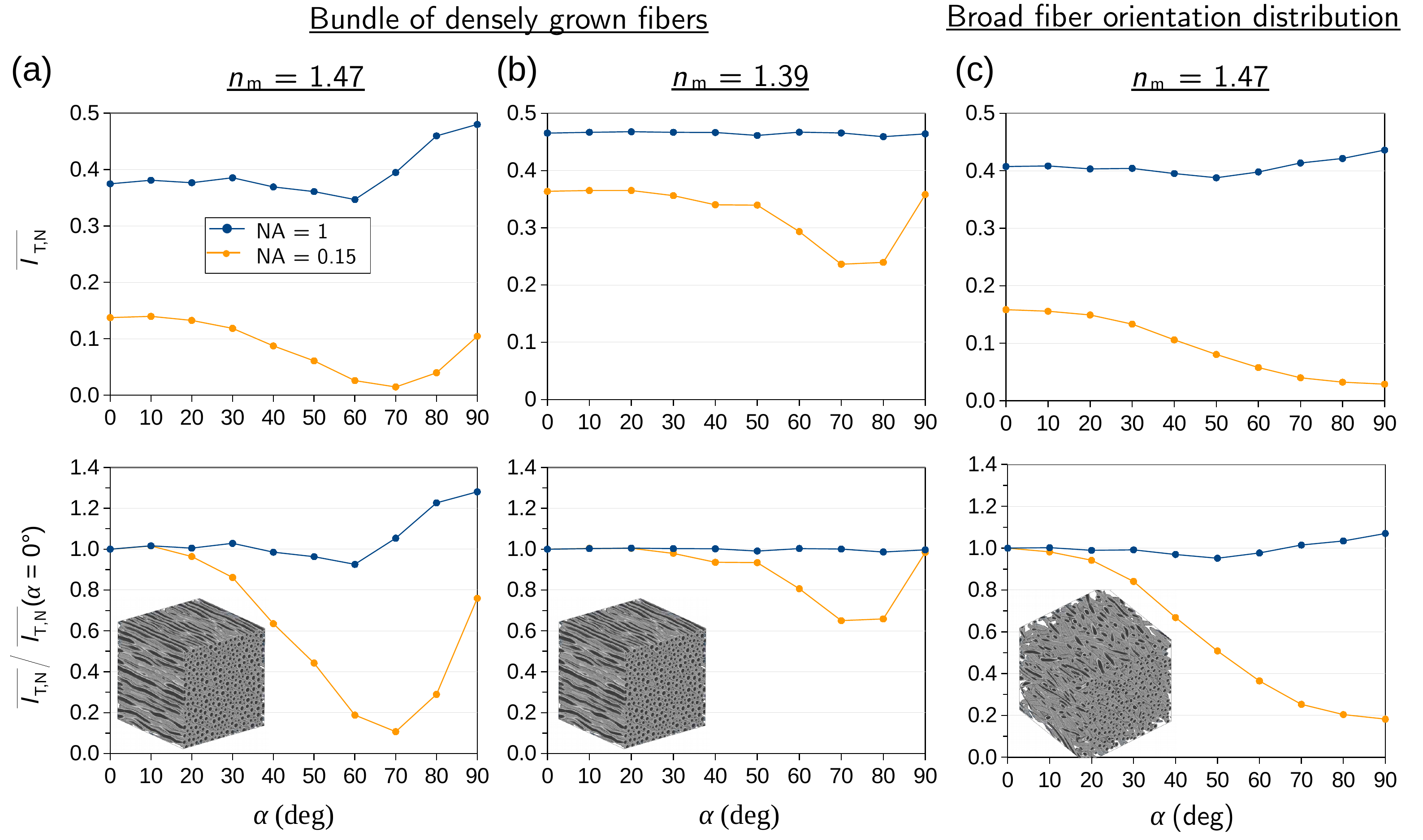}
	\caption{\textbf{Supplementary Figure} $\vert$ Simulated transmittance curves. Mean normalized transmittance values $\overline{I_{\text{T,N}}}$ plotted against the inclination angles $\alpha$ for different fiber bundles and different myelin refractive indices $\nmyelin$: (\textbf{a}) bundle of densely grown fibers with $\nmyelin = 1.47$ (corresponds to given literature values of lipids), (\textbf{b}) bundle of densely grown fibers with $\nmyelin = 1.39$ (the reduced myelin refractive index models tissue with long embedding time), (\textbf{c}) bundle with broad fiber orientation distribution and $\nmyelin = 1.47$.
		The orange curves were computed from 3D-PLI simulations for the numerical aperture of the imaging system (NA = 0.15) and the blue curves for NA = 1.
		To enable a better comparison between horizontal and vertical fibers, the lower figures show the transmittance curves normalized by the mean transmittance of the horizontal fiber bundle, respectively. The 3D-PLI simulations were performed for a normally incident plane wave with 550\,nm wavelength and simulation parameters specified in \cref{tab:Simulation_Parameters}.
		A reduced myelin refractive index (long embedding time) leads to larger transmittance values and a smaller decrease for steep fibers for NA = 0.15. While the transmittance curves for the bundle of densely grown fibers show a minimum for steep fibers ($\alpha = 70$--$80^{\circ}$), the mean transmittance values for the bundle with broad fiber orientation distribution decrease monotonically with increasing fiber inclination.}
	\label{fig:Sim_TransmittanceCurves}
\end{figure*}


\begin{figure*}[!htb]
	\centering
	
	\begin{tikzpicture}[node distance=2.5cm]
	
	\node (TDME3D) [step1, align=center,xshift=-10cm]						
	{\underline{TDME3D} \\[5pt] 
		$\{\vec{A_k}, \,\vec{B}_k, \,\vec{k}\}$
	};
	
	\node (yee) [step2, below of=TDME3D, align=center] 	
	{\underline{Yee Shift (in $j \in \{x,y,z\}$)} \\[2pt] 
		\begin{minipage}{0.65\textwidth}
		\begin{gather*}
		\left. \begin{aligned}
		A_{k,i} \,\cos(k_j \, \Delta j) - B_{k,i} \,\sin(k_j \, \Delta j) \\
		A_{k,i} \,\sin(k_j \, \Delta j) + B_{k,i} \,\cos(k_j \, \Delta j)
		\end{aligned} \,\, \right\}
		\begin{aligned}
		\{\vec{A}'_k,\, \vec{B}'_k,\, \vec{k}\} \quad\quad 
		\end{aligned}
		\end{gather*}
		\end{minipage}
	};
	
	\node (diffraction) [step3, below of=yee, align=center] 	
	{\underline{Scattering Pattern} \\[5pt] 
		$I_k = \frac{\vert \vec{A}'_k \vert^2 + \vert \vec{B}'_k \vert^2}{I_0 / (\text{\#\,px})}$
	};
	
	\node (microlens) [step4, below of=diffraction, align=center] 
	{\underline{Microlenses \& Aperture} \\[5pt]
		$\mathcal{\vec{E}}'_k \equiv \left( \vec{A}'_k + \I \vec{B}'_k \right) \, 2\,\frac{J_1(\kxy \,\cdot\, 0.665\,\um)}{(\kxy \,\cdot\, 0.665\,\um)}$\\
		$\theta_k = \arccos\left(\frac{\kz}{|\vec{k}|}\right) < 8.63^{\circ}$	
	};
	
	\node (FFT_PLI) [step5, below of=microlens, yshift=-1.8cm, align=center] 
	{ 	\underline{2D inverse FFT} \\
		\vspace{0.2cm}
		\begin{minipage}[t][3.8cm]{8.6cm}
		\begin{gather*}
		\left. \begin{aligned}
		\mathcal{E}'_{\text{x}}(\vec{r}) \,&\equiv \iDFT \{ \mathcal{E}'_{k,\text{x}} \} \\
		\mathcal{E}'_{\text{y}}(\vec{r}) \,&\equiv \iDFT \{ \mathcal{E}'_{k,\text{y}} \}
		\end{aligned} \quad\quad\,\,\,\,\right\}
		\begin{aligned}
		&c_0, c_2, d_2 \\
		&\text{(\cref{eq:FourierCoefficients_c_d})}\quad\quad\,\,\,\,
		\end{aligned} \\
		\left. \begin{aligned}
		X_{\text{x}}(\vec{r}) &\equiv \iDFT \{ \mathcal{E}'_{k,\text{x}}\, \kx/\kz \} \\
		X_{\text{y}}(\vec{r}) &\equiv \iDFT \{ \mathcal{E}'_{k,\text{x}}\, \ky/\kz \} \\
		Y_{\text{x}}(\vec{r}) &\equiv \iDFT \{ \mathcal{E}'_{k,\text{y}}\, \kx/\kz \} \\
		Y_{\text{y}}(\vec{r}) &\equiv \iDFT \{ \mathcal{E}'_{k,\text{y}}\, \ky/\kz \}
		\end{aligned} \right\}
		\begin{aligned}
		&e_0, e_2, f_2, e_4, f_4 
		\end{aligned}
		\end{gather*}
		\end{minipage}
	};
	
	\node (I_PLI) [step6, below of=FFT_PLI, yshift=-1.8cm, align=center] 
	{ 	\underline{Normalized Intensity} \\
		\begin{minipage}[t][2.2cm]{8.6cm}
		\begin{align*}
		I_{\text{N}}(\vec{r},\rho) = \,& \big( (c_0 + e_0) + (c_2 + e_2)\,\cos(2\rho) \\
		&+ (d_2 + f_2)\,\sin(2\rho) \\
		&+ e_4 \,\cos(4\rho) + f_4 \,\sin(4\rho) \big) / \big(I_0 / (\text{\#\,px}) \big)
		\end{align*}
		\end{minipage}
	};
	
	\draw [arrow] (TDME3D) -- (yee);
	\draw [arrow] (yee) -- (diffraction);
	\draw [arrow] (diffraction) -- (microlens);
	\draw [arrow] (microlens) -- (FFT_PLI);
	\draw [arrow] (FFT_PLI) -- (I_PLI);
	
	\end{tikzpicture}
	
	\caption{\textbf{Supplementary Figure} $\vert$ Flow chart visualizing the computation of the transmitted light intensities (see \cref{sec:computation-light-intensities}). The electromagnetic field components behind the sample (represented by a set of vectors $\{\vec{A}_k$, $\vec{B}_k$, $\vec{k}\}$, green box) were computed by TDME3D and shifted to the middle of the corresponding Yee cell (with side length $\varDelta$). To study how much light is scattered under a certain angle (wave vector $\vec{k}$), the scattering pattern $I_k$ was computed, \ie, the intensity per wave vector normalized by the ingoing light intensity $I_0$ per image pixel (orange box). The spherical microlenses of the camera detector were modeled by applying a moving average over the area of the microlens with radius 0.665\,\um\ ($J_1$ is the Bessel function of the first kind of order one). The numerical aperture of the imaging system was modeled by considering only waves with directions of propagation $\vec{k}$ that fulfill $\theta_k < \arcsin(\text{NA}) \approx 8.63^{\circ}$. Applying a 2D inverse discrete Fast Fourier Transform (FFT), the transmitted light intensities were computed and normalized by the ingoing light intensity per pixel (red box).}
	\label{fig:FlowChart}
\end{figure*}
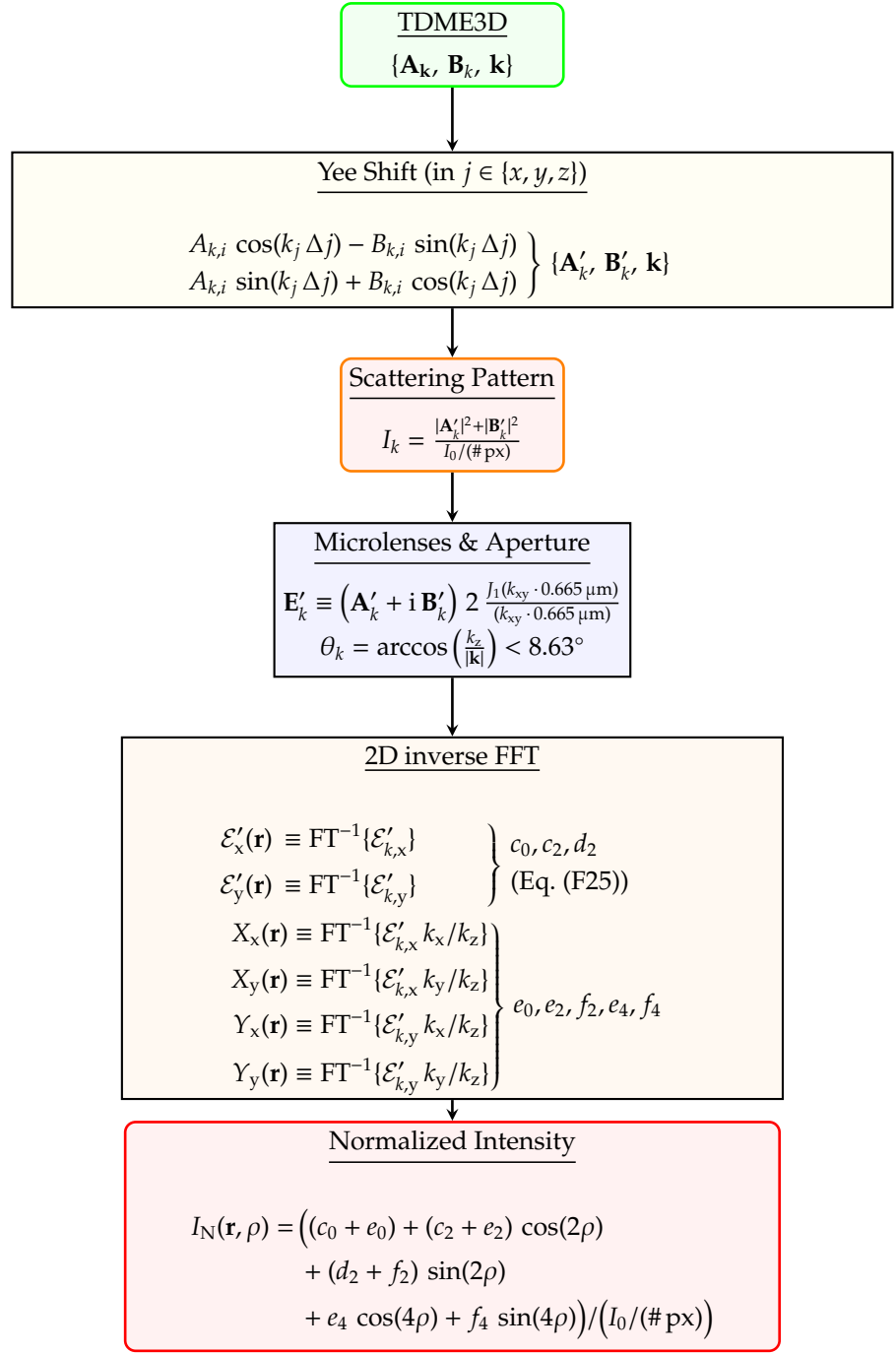


\end{document}